\documentclass[prb,aps,onecolumn,showpacs,floats,a4paper,times,12pt]{revtex4}
\usepackage[dvips]{graphicx}
\usepackage{epsfig,bm,graphics}
\usepackage{epsfig}
\usepackage[francais]{babel}
\usepackage[utf8]{inputenc}
\usepackage{amsmath,amssymb}

\begin{document}
\title{Theory and Simulation of Spin Transport in Antiferromagnetic Semiconductors: Application to MnTe}
\author{K. Akabli$^{a,b}$,  Y. Magnin$^{b}$, Masataka Oko$^{a}$, Isao Harada$^{a}$, and H. T. Diep$^{b}$\footnote{ Corresponding author, E-mail:diep@u-cergy.fr }}
\address{$^{a}$ Graduate School of Natural Science and Technology, Okayama University\\
3-1-1 Tsushima-naka, Kita-ku, Okayama 700-8530, Japan.\\
$^{b}$ Laboratoire de Physique Th\'eorique et Mod\'elisation,
Universit\'e de Cergy-Pontoise, CNRS, UMR 8089\\
2, Avenue Adolphe Chauvin, 95302 Cergy-Pontoise Cedex, France.}
\begin{abstract}
We study in this paper the parallel spin current in an antiferromagnetic semiconductor thin film where we take into account the interaction between itinerant spins and lattice spins. The spin model is an anisotropic Heisenberg model.
We use here the Boltzmann's equation  with numerical data on cluster distribution obtained by Monte Carlo simulations and cluster-construction algorithms. We study the cases of degenerate and non-degenerate semiconductors. The spin resistivity  in both cases is
shown to depend on the temperature with a broad maximum at the transition temperature of the lattice spin system.  The shape of the maximum depends on the spin anisotropy and  on the magnetic field. It shows  however no sharp peak in contrast to ferromagnetic materials.
Our method is applied to  MnTe. Comparison to experimental data is given.
\end{abstract}
\maketitle

\section{Introduction}

 The behavior of the spin resistivity $\rho$
 as a function of temperature ($T$) has been shown and theoretically explained by many authors during the last 50 years. Among the ingredients which govern the properties of $\rho$, we can mention
 the scattering of the itinerant spins by the lattice magnons suggested by Kasuya\cite{Kasuya},
 the diffusion due to impurities\cite{Zarand}, and
 the spin-spin correlation.\cite{DeGennes,Fisher,Kataoka}  First-principles analysis of spin-disorder resistivity of Fe and Ni has been also recently performed.\cite{Wysocki}

 Experiments have been performed on  many magnetic materials ranging from metals to semiconductors.   These results show that the behavior of the spin resistivity depends on the material: some of them show a large peak of $\rho$ at the magnetic transition temperature $T_C$,\cite{Matsukura} others show only a change of slope of $\rho$ giving rise to a peak of the differential resistivity $d\rho/dT$.\cite{Stishov,Shwerer}  Very recent experiments such as those performed on ferromagnetic SrRuO$_3$ thin films\cite{Xia}, Ru-doped induced ferromagnetic La$_{0.4}$Ca$_{0.6}$MnO$_3$\cite{Lu},  antiferromagnetic $\epsilon$-(Mn$_{1-x}$Fe$_x$)$_{3.25}$Ge\cite{Du}, semiconducting Pr$_{0.7}$Ca$_{0.3}$MnO$_3$ thin films\cite{Zhang},superconducting BaFe$_2$As$_2$ single crystals\cite{Wang-Chen}, La$_{1-x}$Sr$_x$MnO$_3$\cite{Santos}  and Mn$_{1-x}$Cr$_x$Te\cite{Li} compounds show different forms of anomaly of the magnetic resistivity at the magnetic phase transition temperature.

 The magnetic resistivity due to the scattering of itinerant spins by localized lattice spins is proportional to the spin-spin correlation as proposed long-time ago by De
Gennes and Friedel\cite{DeGennes}, Fisher and Langer\cite{Fisher}, and recently by
Kataoka\cite{Kataoka}.   They have shown that changing the range of spin-spin correlation changes  the shape of $\rho$.
In a recent work, Zarand  et al.\cite{Zarand} have showed that in magnetic diluted semiconductors the shape of the resistivity
versus $T$ depends on the interaction between the itinerant spins and
localized magnetic impurities which is characterized by a Anderson localization length $\zeta$.  Expressing physical quantities in terms of $\zeta$ around impurities, they calculated $\rho$ and showed that its peak height depends indeed on this localization length.

 In our previous work\cite{Akabli,Akabli2,Akabli3} we have studied the spin current in ferromagnetic thin films. The behavior of the spin resistivity
 as a function of  $T$ has been shown and explained as an effect of magnetic domains formed in the proximity of the phase transition point.  This new concept has an advantage over the use of the spin-spin correlation since the distribution of clusters is more easily calculated using Monte Carlo simulations.
 Although the formation of spin clusters and their sizes are a consequence of spin-spin correlation,
the direct access in numerical calculations to the structure of clusters allows us to study complicated systems
 such as thin films, systems with impurities, systems with high degree of instability etc. On the other hand, the  correlation
functions are very difficult to calculate. Moreover, as will be shown in this paper, the correlation function
cannot be used  to explain the behavior of the spin resistivity in antiferromagnets where very few theoretical
investigations have been carried out. One of these is the work by Suezaki and Mori\cite{Mori} which simply
 predicted that the behavior of the spin resistivity in antiferromagnets is that in ferromagnets if the
correlation is short-ranged.   It means that correlation should be limited to "selected nearest-neighbors". Such an explanation is obviously not satisfactory in particular when the sign of the correlation function between
antiparallel spin pairs are taken into account.  In a work with a model suitable for magnetic semiconductors,
 Haas has shown that the resistivity $\rho$  in antiferromagnets is quite different from that of ferromagnets.\cite{Haas} In particular, he found  that while ferromagnets show a peak of $\rho$ at the magnetic transition of the lattice spins, antiferromagnets do not have such a peak.   We will demonstrate that all these effects can be interpreted in terms of clusters used in our model.

In this paper, we introduce a simple model which takes into account  the interaction between itinerant spins and localized lattice spins.   This is similar to the $s-d$ model\cite{Haas}.  The lattice spins interact with each other via antiferromagnetic interactions.   The model will be studied  here by a combination of Monte Carlo simulation and the Boltzmann's equation.   As will be discussed below, such a model corresponds to antiferromagnetic semiconductors such as MnTe.  An application is made for this compound in the present work.

The paper is organized as follows. In section \ref{theo}, we show and discuss
our general model and its application to the antiferromagnetic case using the Boltzmann's
equation formulated in terms of clusters.  We  also describe here our Monte Carlo simulations to obtain the distributions of sizes and number of clusters as functions of $T$ which will be used to solve the  Boltzmann's
equation.  Results on the effects of Ising-like anisotropy and magnetic field as well as an application to the case of MnTe is shown in section III.
Concluding remarks are given in section IV.

\section{Theory}\label{theo}

Let us recall briefly principal theoretical models for magnetic
resistivity $\rho$. In magnetic systems, de Gennes and Friedel\cite{DeGennes} have
suggested that the magnetic resistivity is proportional to the spin-spin
correlation.  As a consequence, in ferromagnetically ordered systems,
$\rho$ shows a divergence
at the transition temperature $T_C$, similar to the susceptibility.  However, in order to explain the finite
cusp of $\rho$ experimentally observed in some experiments, Fisher
and Langer\cite{Fisher} suggested to take into account only
short-range correlations in the de Gennes-Friedel's theory.
Kataoka\cite{Kataoka} has followed the same line in proposing a
model where he included, in addition to a parameter describing the correlation range,
some other parameters describing effects of the magnetic  instability, the
density of itinerant spins and the applied magnetic field.

For  antiferromagnetic systems,  Suezaki and Mori\cite{Mori} proposed a model to explain the anomalous behavior of the resistivity around the
N\'eel temperature. They used the Kubo's formula  for an  $s-d$ Hamiltonian with some approximations to connect the resistivity
to the correlation function. However, it is not so easy to resolve the problem. Therefore, the form of the correlation function was just given in the molecular field approximation. They argued that just below the N\'eel temperature $T_N$ a long-range correlation appears giving rise to an additional magnetic potential which causes a gap. This gap affects the electron density which alters the spin resistivity but does not in their approximation interfere in the scattering mechanism.  They concluded that, under some considerations, the resistivity should have
a peak close to the N\'eel point. This behavior is observed in $Cr$, $\alpha-Mn$ and some rare earth metals.  Note however that in the approximations used by Haas\cite{Haas}, there is no peak predicted.  So the question of the existence of a peak in antiferromagnets remains open.

Following Haas, we use for semiconductors the following interaction
\begin{equation}
V=\sum_nJ(\vec r-\vec R_n)\mathbf  s\cdot \mathbf S_n \label{int}
\end{equation}
where $J(\vec r-\vec R_n)$ is the exchange interaction between an itinerant
spin $\mathbf s$ at $\vec r$ and the lattice spin $\mathbf S_n$ at the lattice site $\vec R_n$.
In practice, the sum on lattice spins $\mathbf S_n$
should be limited at some cut-off distance as will be discussed later.
Haas supposed that $V$ is weak enough to be
considered as a perturbation to the lattice Hamiltonian given by Eq. (\ref{eqn:hamil}) below. This is what we also suppose in the present paper.
He applied his model to ferromagnetic doped CdCr$_2$Se$_4$\cite{Lehmann,Shapira,Snyder} and antiferromagnetic
semiconductors MnTe.  Note however that the model by Haas as well as other existing models cannot treat the case
where itinerant spins, due to the interaction between themselves, induce itinerant magnetic ordering such as in
(Ga,Mn)As shown by Matsukura et al.\cite{Matsukura}
Note also that both the up-spin  and down-spin currents are present in the theory but the authors considered only the
effect of the up-spin current since the interaction "itinerant spin"-"lattice spin" is ferromagnetic so that the
down-spin current is very small.  This theory was built in the framework of the relaxation-time approximation of
the Boltzmann's equation under an electric field. As De Gennes and Friedel, Haas used here the spin-spin correlation
to describe the scattering of itinerant spins by the disorder of the lattice spins. As a result, the model of Haas
shows a peak in the ferromagnetic case but no peak in the antiferromagnetic semiconductors.  Experimentally, the
absence of a peak has been observed in antiferromagnetic  LaFeAsO by McGuire et al.\cite{McGuire}
and in CeRhIn$_5$ by Christianson et al.\cite{Christianson}

\subsection{Boltzmann's equation}

In the case of Ising spins in a ferromagnet that we studied before\cite{Akabli3}, we have
made a theory based on the cluster structure of the lattice spins. The cluster  distribution was incorporated in the
Boltzmann's equation.  The number of clusters $\eta$ and their sizes $\xi$ have been
numerically determined using the Hoshen-Kopelmann's algorithm (section~\ref{HoshenKopelmann}).\cite{Hoshen}
We work in diffusive regime with approximation of parabolic band and in an $s-d$ model.
 We consider in this paper that in our range of temperature the Hall resistivity is constant (constant density).
To work with the Born approximation we consider a weak potential of interaction between clusters of spin and conduction electrons. We
suppose that the life's time of clusters is larger than the relaxation time.
 As in our previous paper\cite{Akabli3} we use in this paper the expression of relaxation time obtained from the Boltzmann's equation in the following manner. We first write the Boltzmann's equation for $f$, the
distribution function of itinerant electrons, in a uniform electric field $\textbf E$
\begin{equation}
(\dfrac{\hbar \textbf{k}.e\textbf{E}}{m})(\dfrac{\partial
f^0}{\partial\varepsilon})=(\dfrac{\partial f}{\partial
t})_{coll}, \label{eqn:equ10}
\end{equation}
where $f^0$ is the equilibrium Fermi-Dirac function, $\textbf{k}$ the wave vector, $e$ and $m$ the electronic
charge and mass, $\epsilon$ the electron energy.  We next use the
following relaxation-time approximation
\begin{equation}
(\dfrac{\partial f_k}{\partial t})_{coll}= -
(\dfrac{f_k^{1}}{\tau_k}), \qquad f_k^{1}=f_k-f_k^{0},
\label{eqn:equ20}
\end{equation}
where $\tau_k$ is the relaxation time. Supposing  elastic
collisions, i. e. $k=k'$,  and using the detailed balance we have
\begin{equation}
(\dfrac{\partial f_k}{\partial
t})_{coll}=\dfrac{\Omega}{(2\pi)^{3}}\int
[w_{k',k}(f^{1}_{k'}-f^{1}_{k})] d\mathbf {k'}, \label{eqn:equ30}
\end{equation}
where $\Omega$ is the system volume, $w_{k',k}$ the transition
probability between $\textbf k$ and $\textbf k'$. We find with
Eq.~(\ref{eqn:equ20}) and Eq.~(\ref{eqn:equ30}) the following
well-known  expression
\begin{eqnarray}
(\dfrac{1}{\tau_k})&=&\dfrac{\Omega}{(2\pi)^{3}}\int
[w_{k',k}(1-\cos\theta)] \nonumber\\
&&\times \sin\theta k'^{2}dk' d\theta d\phi, \label{eqn:equ40}
\end{eqnarray}
 where $\theta$ and $\phi$ are the angles formed by $\textbf k'$ with $\textbf k$, i. e. spherical
 coordinates with $z$ axis parallel to $\mathbf {k}$.

We use now in Eq.~(\ref{eqn:equ40}) the "Fermi golden rule" for $\omega_{k,k'}$ and we
obtain

\begin{subequations}
  \begin{gather}
\dfrac{1}{\tau_k}=\dfrac{\Omega}{(2\pi)^3}\int [\omega_{k,k'}(1-cos(\theta))]sin(\theta)k'^2 dk' d\theta d\phi\\
\omega_{k,k'}=\dfrac{(2\pi)m}{\hbar^3 k} |<k'|J(r)|k>|^2 \delta(k'-k)
  \end{gather}
\end{subequations}
where $J(r)$ is the exchange integral between an itinerant spin and a lattice spin which is given in the scattering potential, Eq. (\ref{int}). One has
\begin{equation}
J(r)\equiv J(|\vec r'-\vec R_n|)
\end{equation}
Note that for simplicity we have supposed here that the interaction potential $J(r)$ depends only on the relative distance $r'=|\vec r-\vec R_n|$, not on the direction of $\vec r-\vec R_n$.  We suppose in the following a potential which exponentially decays with distance
\begin{equation}\label{exchange}
J(r)\equiv V_0e^{-r/\xi}
\end{equation}
where $V_0$ expresses the magnitude of the interaction and $\xi$ the averaged cluster size.   After some algebra, we arrive at the following relaxation time
\begin{equation}
\dfrac{1}{\tau_{k_f}}=\dfrac{32V_0^2 m \pi}{(2k\hbar)^3}\eta \xi^2[1-\dfrac{1}{1+(2\xi k_f)^2}-\dfrac{(2\xi k_f)^2}{[1+(2\xi k_f)^2]^2}]
\end{equation}
where $k_f$ is the Fermi wave vector.
As noted by Haas\cite{Haas}, the mobility is inversely proportional to the susceptibility $\chi$. So,
in examining our expression and in using the following expression $\chi=\sum \xi^2 \eta(\xi)$,\cite{Binder} where $\eta (\xi)$ is the number of clusters of size $\xi$,
one sees that the first term of the relaxation time is proportional to the susceptibility. The other two
terms are the corrections.

The mobility in the $x$ direction is defined by
\begin{equation}
\mu_x=\dfrac{e\hbar^2}{3m^2}\dfrac{\sum_k k^2(\partial f^0_k/\partial \epsilon)\tau_k}{\sum_k f^0_k}
\end{equation}
We resolve the mobility $\mu_x$ explicitly in the following two cases
\begin{itemize}
\item \underline{Degenerate semiconductors}

\begin{subequations}
  \begin{gather}
\sum_k f^0_k=2\pi (\dfrac{2m}{\hbar^2})^{3/2}[\dfrac{2}{3} \epsilon_f^{3/2}]\\
\sum_k k^2(\partial f^0_k/\partial \epsilon)\tau_k =2 \pi (\dfrac{2m}{\hbar^2})^{3/2} \dfrac{\epsilon_f^{1/2}}{D} (\dfrac{2m\epsilon_f}{\hbar^2})^{5/2} [\dfrac{1+8m\xi^2\epsilon_f/\hbar^2}{8m\xi^2\epsilon_f/\hbar^2}]^2
  \end{gather}
\end{subequations}
where $D=\dfrac{\eta 4V_0^2 m \pi \xi^2}{\hbar^3}$. We arrive at the following mobility
\begin{subequations}
  \begin{gather}
\mu_x=\dfrac{e\hbar^2}{2m^2} \dfrac{\epsilon_f^{-1}}{D} (\dfrac{2m\epsilon_f}{\hbar^2})^{5/2} [\dfrac{1+8m\xi^2\epsilon_f/\hbar^2}{8m\xi^2\epsilon_f/\hbar^2}]^2 \\
\sigma=n e \mu = \dfrac{n e^2}{mDk_f}[\dfrac{1+4\xi^2 k^2_f}{4\xi^2}]^2
  \end{gather}
\end{subequations}
The resistivity is then
\begin{subequations}
  \begin{gather}
\rho= \dfrac{\eta 4V_0^2 m^2 \pi k_f \xi^2}{n e^2 \hbar^3}[\dfrac{4\xi^2}{1+4\xi^2 k^2_f}]^2
  \end{gather}
\end{subequations}
We can check  that the right-hand side has the dimension of a resistivity: $\dfrac{[kg][m]^3}{[C]^2[s]}=[\Omega][m]$.

\item \underline{Non-degenerate semiconductors}

One has in this case $f_k^0=exp(-\beta \epsilon_k)$\\
\begin{subequations}
  \begin{gather}
\sum_k f^0_k=2\pi (\dfrac{2m}{\hbar^2})^{3/2} \beta^{-3/2} \sqrt{\pi}/2 \\
\sum_k k^2(\partial f^0_k/\partial \epsilon)\tau_k = 2\pi (\dfrac{2m}{\hbar^2})^{3/2} \dfrac{1}{2D(4\xi^2)^2\beta} (\dfrac{2m}{\hbar^2})^{1/2}[1+\dfrac{2\times 16m\xi^2}{\hbar^2 \beta} +\dfrac{6(8m\xi^2)^2}{\hbar^4 \beta^2}] \\
\sigma = ne\mu=  \dfrac{ne^2\hbar^2}{m^2D(4\xi^2)^2\sqrt{\pi}} (\dfrac{2m\beta}{\hbar^2})^{1/2}[1+\dfrac{2\times 16m\xi^2}{\hbar^2 \beta} +\dfrac{6(8m\xi^2)^2}{\hbar^4 \beta^2}]\\
\rho=\dfrac{1}{\sigma}
  \end{gather}
\end{subequations}
where $D=\dfrac{\eta 4V_0^2 m \pi \xi^2}{\hbar^3}$
\end{itemize}
Note that the formulation of our theory in terms of cluster number $\eta$ and cluster size $\xi$ is numerically very
convenient.  These quantities are easily calculated by Monte Carlo simulation for the Ising model. The method can be generalized
to the case of Heisenberg spins where the calculation  is more
complicated as seen below.  In section~\ref{E_Ising} we will examine values of parameter $V_0$
where the Born's approximation is valid.

\subsection{Algorithm of Hoshen-Kopelmann and Wolff's procedure}\label{HoshenKopelmann}

We use the  Heisenberg spin model with an Ising-like anisotropy for an
antiferromagnetic film of body-centered cubic (BCC) lattice of  $N_x\times N_y \times N_z$
cells where there are two atoms per cell. The film has two symmetrical (001) surfaces, i.e. surfaces perpendicular to the $z$ direction.
We use the periodic boundary conditions in
the $xy$ plane and the mirror reflections in the $z$ direction.  The lattice
Hamiltonian is written as follows
\begin{equation}
\mathcal H= J\sum_{\left<i,j\right>}\mathbf S_i\cdot\mathbf S_j +A\sum_{\left<i,j\right>} S^z_i S^z_j
\label{eqn:hamil}
\end{equation}
where $\mathbf S_i$ is the Heisenberg spin at the site $i$,
$\sum_{\left<i,j\right>}$ is performed over all nearest-neighbor (NN) spin pairs. We assume here that all interactions including
those at the two surfaces are identical for simplicity:
$J$ is positive (antiferromagnetic), and $A$ an Ising-like anisotropy which is a positive constant.
When $A$ is zero, one has the isotropic Heisenberg model and when
$A \rightarrow \infty$, one has the Ising model.
The classical Heisenberg spin model is continuous, so it allows the
domain walls to be less abrupt and therefore softens the behavior of the magnetic
resistance.   Note that for clarity of illustration, in this section (II.B) we suppose only NN interaction $J$.  In the application to MnTe shown in section III.C, the exchange integral is distance-dependent and we shall take into account up to the third NN interaction.\\

Hereafter, the temperature is expressed in unit of $J/k_B$, $k_B$ being the Boltzmann's constant. $A$ is given in unit of $J$.  The resistivity $\rho$ is shown in atomic units.

\begin{figure}[!h]
\centering
 \includegraphics[width=120mm]{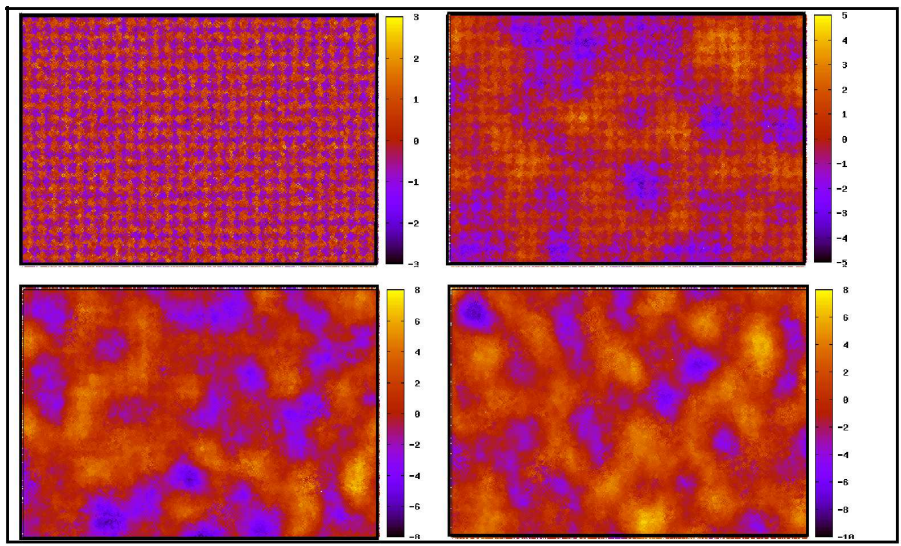}
\caption{Energy map of an itinerant spin in the $xy$ plane with $D_1=2$ in unit of the lattice constant $a$ and $A=0.01$, for $T=0.01$, $T=1.0$, $T=2.0$ and $T=2.5$ (from left to right, top to bottom, respectively).  The values of energy corresponding to different colors are given on the right.}\label{fig:Picture(T)}
 \end{figure}

For the whole paper, we use $N_x=N_y=20$, and $N_z=8$. The finite-size effect as well as surface effects are out of the scope of the present paper.
Using the Hamiltonian (\ref{eqn:hamil}), we equilibrate the lattice at a temperature $T$ by the standard Monte Carlo simulation.
 In order to analyze the spin resistivity, we should know the energy landscape seen by an itinerant spin.
The energy map of an itinerant electron in the lattice is obtained as follows: at each position its energy is calculated using
Eq. (\ref{exchange}) within a cutoff at a distance $D_1=2$ in unit of the lattice constant $a$.
The energy value is coded by a color as shown in Fig.~\ref{fig:Picture(T)} for the case $A=0.01$.
As seen,
at very low $T$ ($T=0.01$) the energy map is periodic just as the lattice,
 i. e. no disorder. At $T=1$, well below the N\'eel temperature $T_N\simeq 2.3$, we observe an energy
map which indicates the existence of many large defect clusters of high energy in the
lattice. For $T \approx T_N$ the lattice is completely disordered. The same is true
for $T=2.5$ above $T_N$.

We shall now calculate the number of clusters and their sizes as a
function of $T$ in order to analyze the temperature-dependent
behavior of the spin current.

\begin{figure}
 \centering
 \includegraphics[width=70mm,angle=-90]{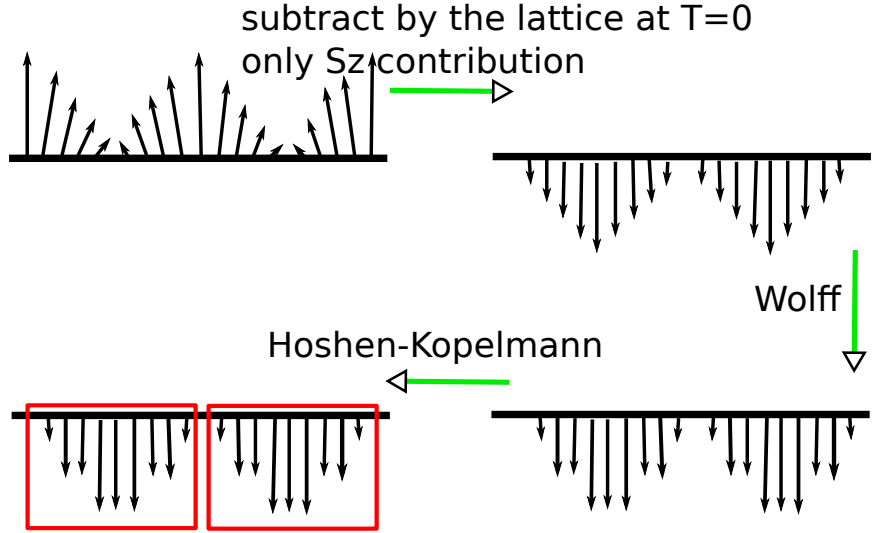}
 \caption{The successive steps in the application of the algorithm by Wolff to the case of Heisenberg spin. See text for explanation.} \label{fig}
\end{figure}

The scattering by clusters in the Ising case in our previous model\cite{Akabli3} is now replaced in the Heisenberg spin model studied here, by a
scattering due to large domain walls. Counting the number of clusters in the
Heisenberg case requires some particular attention as seen in the following:
\begin{itemize}
 \item we equilibrate the system at $T$
 \item we generate  first bonds according to the algorithm by Wolff:\cite{Wolff2,Wolff} it consists in
replacing the two  spins where the link is verified the Wolff's probability, by their larger value
(Fig~\ref{fig})
 \item we next discretize $S_z$, the $z$ component  of each spin, into values between $-1$ and $1$ with a step $0.1$
 \item only then we can use  the algorithm of Hoshen-Kopelmann to form a cluster with the neighboring
spins of the same $S_z$. This is how our clusters in the Heisenberg case are obtained.
\end{itemize}

\begin{figure}[h!]
 \centering
 \includegraphics[width=70mm,angle=-90]{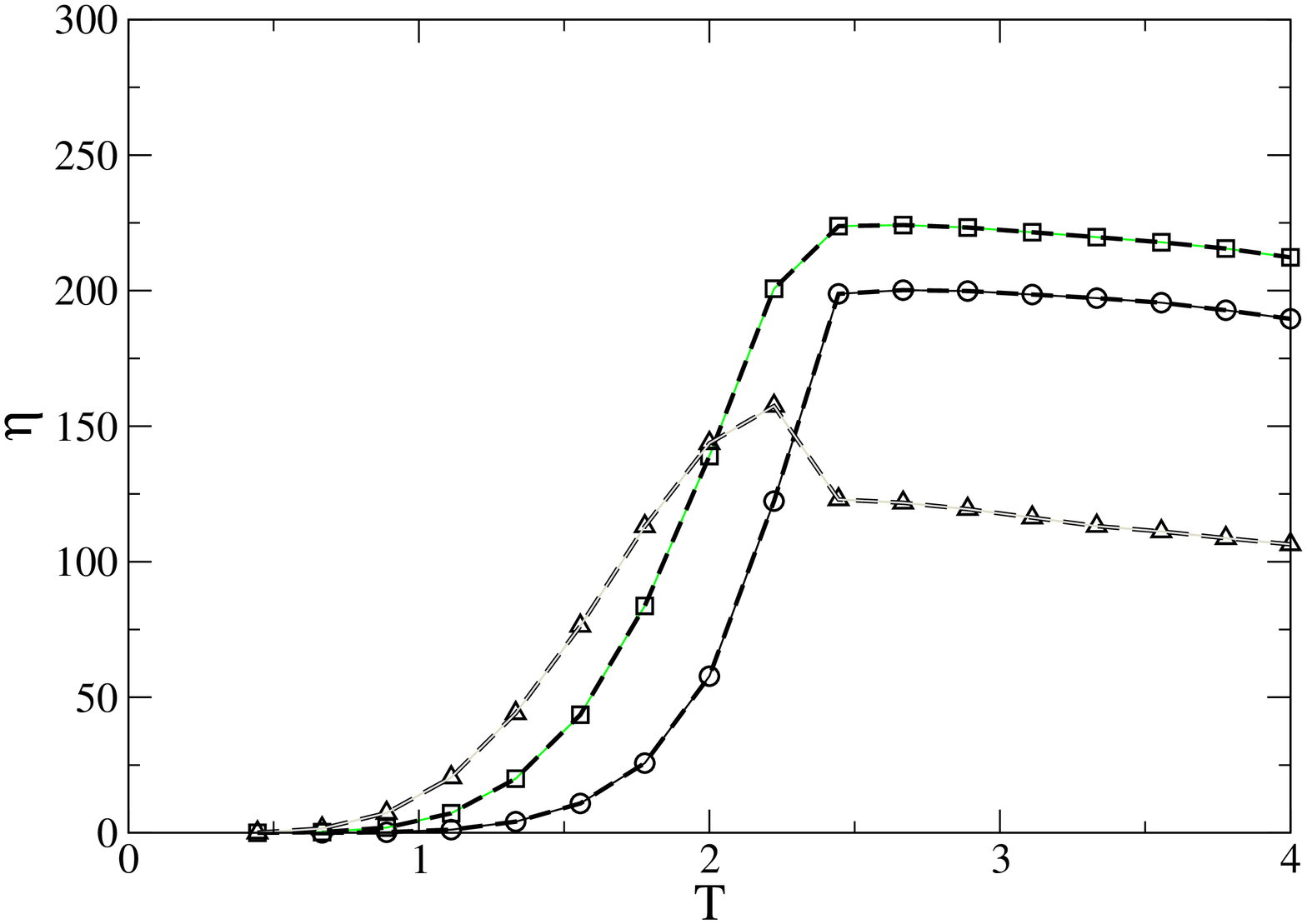}
 \includegraphics[width=70mm,angle=-90]{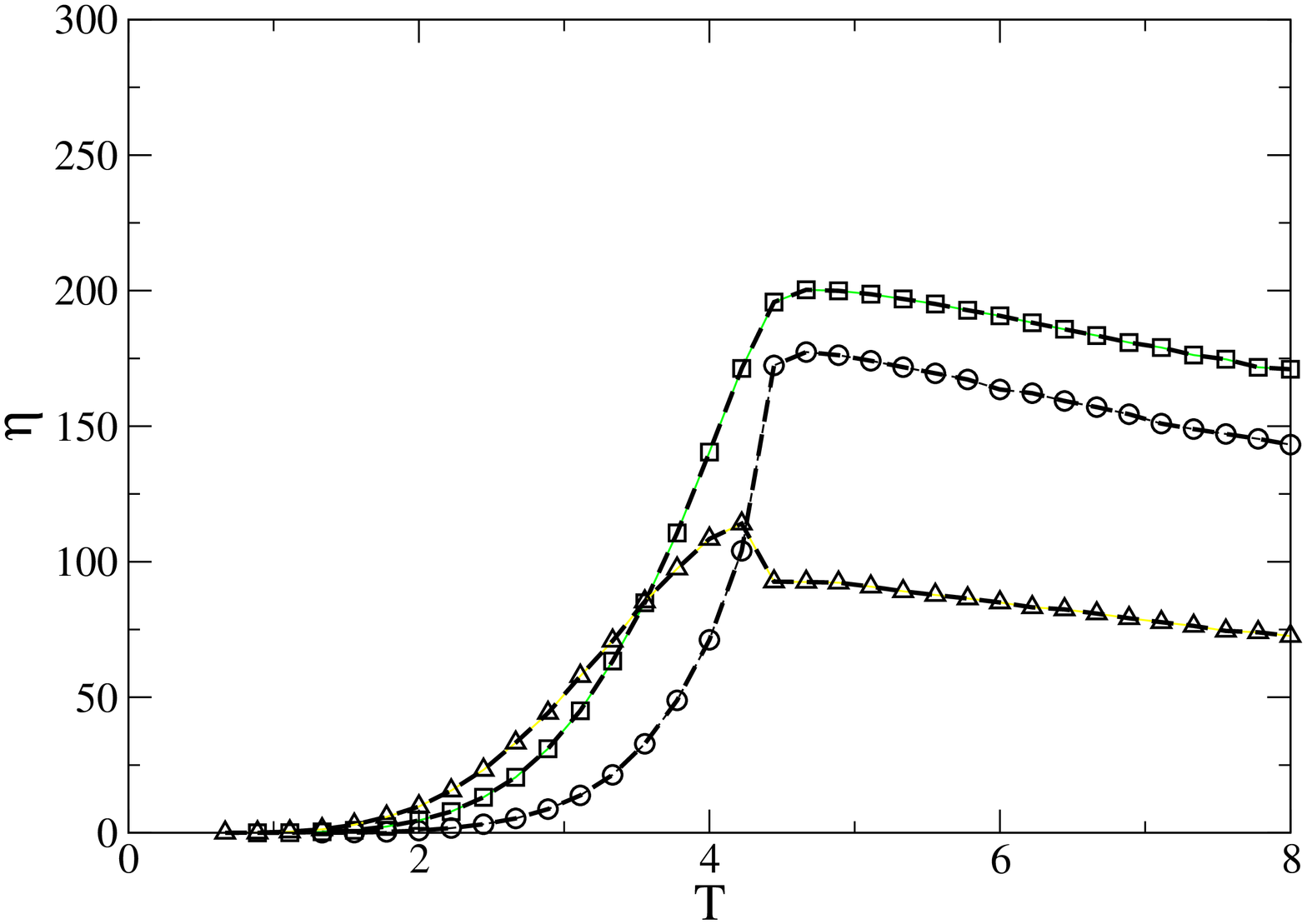}
\caption{Number of clusters versus temperature for anisotropy $A=0.01$ (upper) and  $A=1$ (lower). The values of $S_z$ are 1, 0.8 and 0.6  denoted by circles, squares and triangles, respectively.  Lines are guides to the eye.} \label{histo}
\end{figure}

Note that we can define a cluster distribution by each value of $S_z$. We can therefore distinguish
the amplitude  of scattering: as seen below scattering is stronger for cluster with larger $S_z$.
We have used the above procedure to count the number of clusters
in our  simulation of an antiferromagnetic
thin film. We show in Fig. \ref{histo} the number of cluster $\eta$ versus $T$ for several values of $S_z$.

\begin{figure}[h!]
 \centering
 \includegraphics[width=70mm,angle=-90]{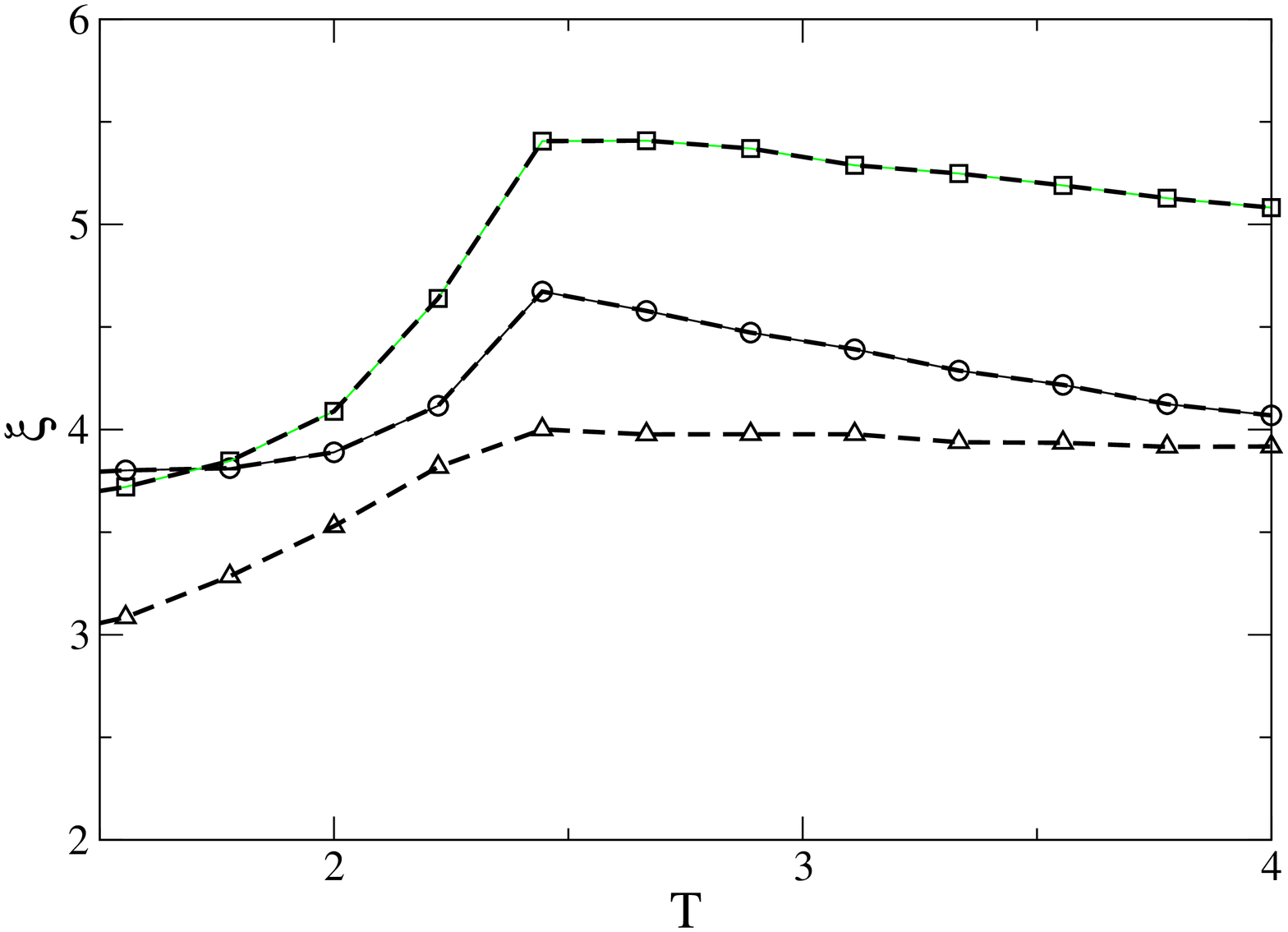}
 \includegraphics[width=70mm,angle=-90]{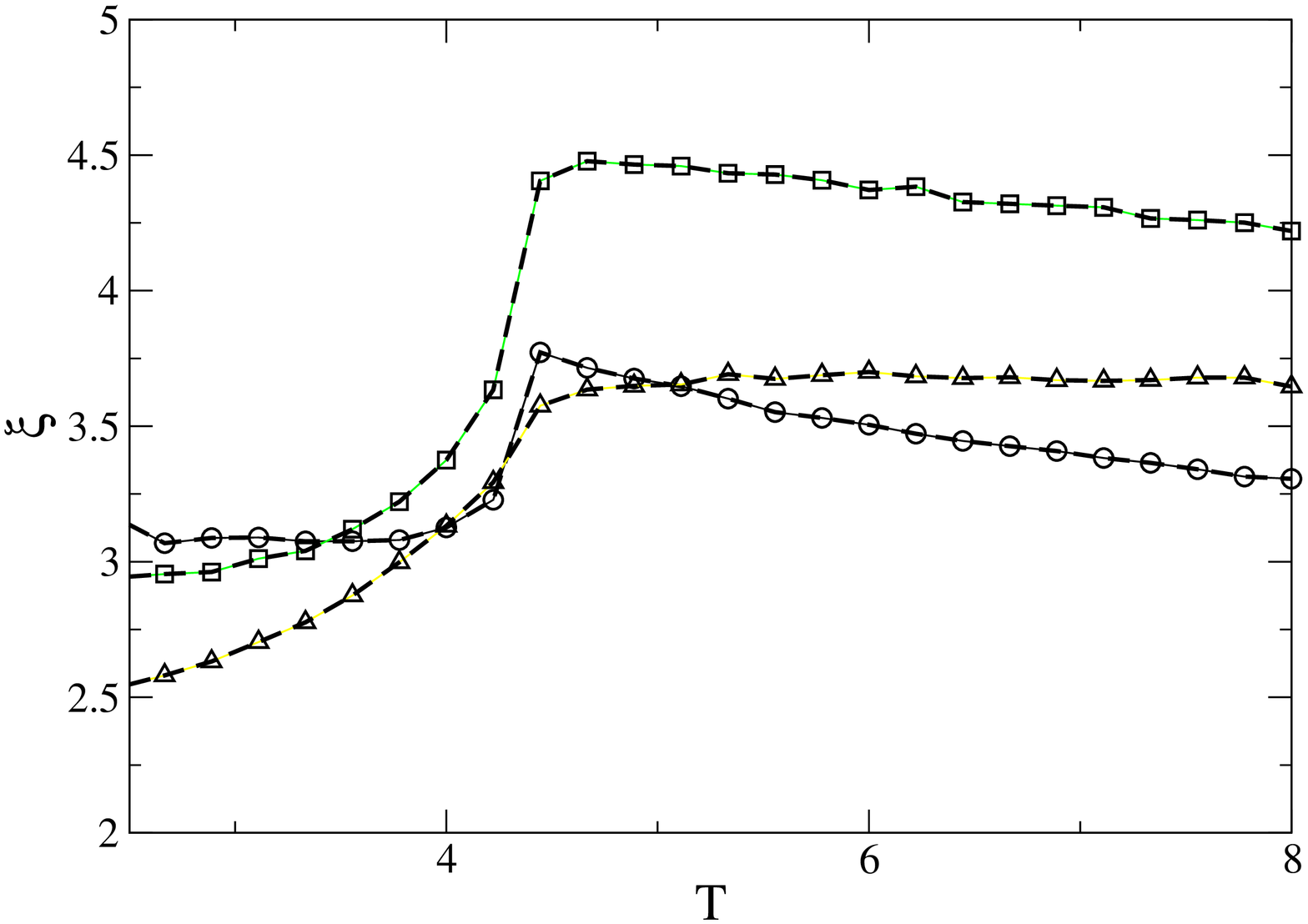}
 \caption{Average size of clusters versus temperature for  anisotropy $A=0.01$
 (upper) and $A=1$ (lower).  The values of $S_z$ are 1, 0.8 and 0.6  denoted by circles, squares and triangles, respectively.  Lines are guides to the eye.} \label{cluster}
\end{figure}

We have in addition determined the average size
of these clusters as a function
of $T$. The results are shown in Fig.~\ref{cluster}.  One observes  that
the size and the number of clusters of any value of $S_z$ change the behavior showing a maximum at the transition temperature.

The resistivity, as mentioned above, depends indeed on the amplitude of $S_z$ as seen in the expression
\begin{equation}
\rho=\dfrac{m}{ne^2}\dfrac{1}{\tau}=\dfrac{m}{ne^2} \sum^{S_z}_{i=-S_z} \dfrac{1}{\tau_i}
\end{equation}

\section{Results}

\subsection{Effect of Ising-like Anisotropy}\label{E_Ising}
\begin{figure}[h!]
 \centering
 \includegraphics[width=55mm,angle=-90]{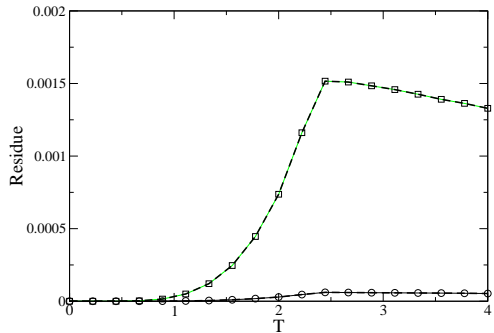}
 \caption{Ratio Residue=$\rho (Born2)/\rho (Born1)$ versus $T$  for  $V_0$=0.05 (squares, upper curve) and 0.01 (circles, lower curve). See text for comments.} \label{fig:Residue}
\end{figure}

At this stage, it is worth to return to examine some fundamental effects of $V_0$ and $A$.
It is necessary to know acceptable values of $V_0$ imposed by the Born's approximation. To do this
we must calculate the resistivity with the second order Born's approximation.
\begin{subequations}
  \begin{gather}
\sigma_k^B(\theta,\phi)=|\dfrac{F(\theta,\phi)}{4\pi}|^2\\
F(\theta,\phi)=\dfrac{2m\Omega}{\hbar^2} [\int d^3r e^{-i{\bf K}.{\bf r}} J(r) - \dfrac{1}{4\pi}\int d^3r e^{-i{\bf K}.{\bf r}} \dfrac{J(r)}{r} \int d^3r' e^{-i{\bf K}.{\bf r'}} J(r')]\\
K=|{\bf k}-{\bf k}'|=k[2(1-\cos\theta)]^{1/2}\ \mbox{and} \ J(r)=V_0e^{-r/\xi}\nonumber
  \end{gather}
\end{subequations}
we find, with $D=\dfrac{\eta 32 \pi \Omega m}{\hbar^3}$,
\begin{equation}
\dfrac{1}{\tau_k}=DV_0^2k[\dfrac{2\xi^6}{[1+(2\xi k)^2]^2} - \dfrac{V_0}{3[1+(2\xi k)^2]^2}(1+\dfrac{4}{[1+(2\xi k)^2]^2}) + \dfrac{V_0^2\xi^6}{12(2k^2)^2}]
\end{equation}
The first term is due to the first order of Born's approximation and the second and third terms to corrections from the second order. We plot $\rho (Born2)/\rho (Born1)$  versus  $T$ in
Fig.~\ref{fig:Residue}  for different values of $V_0$,  $\rho (Born1)$  and $\rho (Born2)$ being respectively the resistivities calculated at the first and second order.  We note that the larger this ratio is, the more important the corrections due to the second-order become.   From Fig.~\ref{fig:Residue}, several remarks are in order:
\begin{itemize}
\item The first
order of Born's approximation is valid for small values of $V_0$ as seen in the case $V_0= 0.01$ corresponding to a few meV.  In this case the resistivity does not depend on $T$. This is understandable because with such a weak coupling to the lattice, itinerant spins do not feel the effect of the lattice spin disordering.
\item  In the case of strong $V_0$ such as $V_0=0.05$, the second-order approximation should be used. Interesting enough, the resistivity is strongly affected by $T$ with a peak corresponding to the phase transition temperature of the lattice.
\end{itemize}

\begin{figure}[h!]
\centering
\includegraphics[width=55mm,angle=-90]{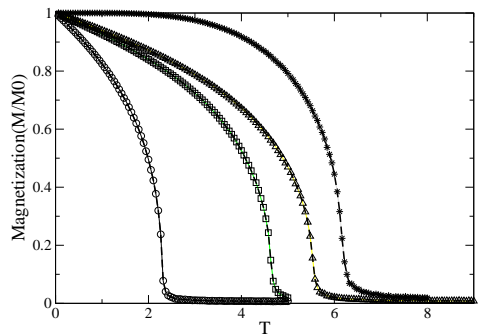}
\caption{Sublattice magnetization versus
 $T$ for several values of anisotropy $A$. From left to right $A=0.01$, $A=1$, $A=1.5$ and pure Ising spin. }\label{fig:MA}
\end{figure}

We examine now the effect $A$.
Figure~\ref{fig:MA} shows  the variation of the sublattice magnetization  and of $T_N$ with anisotropy $A$.
We have obtained respectively for $A=0.01$, $A=1$, $A=1.5$ and pure Ising case the following critical temperatures
$T_N\simeq 2.3$, $4.6$, $5.6$ and $6.0$.  Note that the pure Ising case has been simulated with the pure Ising Hamiltonian, not with Eq. (\ref{eqn:hamil}) (we cannot use $A=\infty$).
We can easily understand that not only the spin resistivity will follow  this variation of $T_N$ but also
the change of $A$ will fundamentally alter the resistivity behavior as will be seen below.\\

\begin{figure}[h!]
 \centering
 \includegraphics[width=70mm,angle=-90]{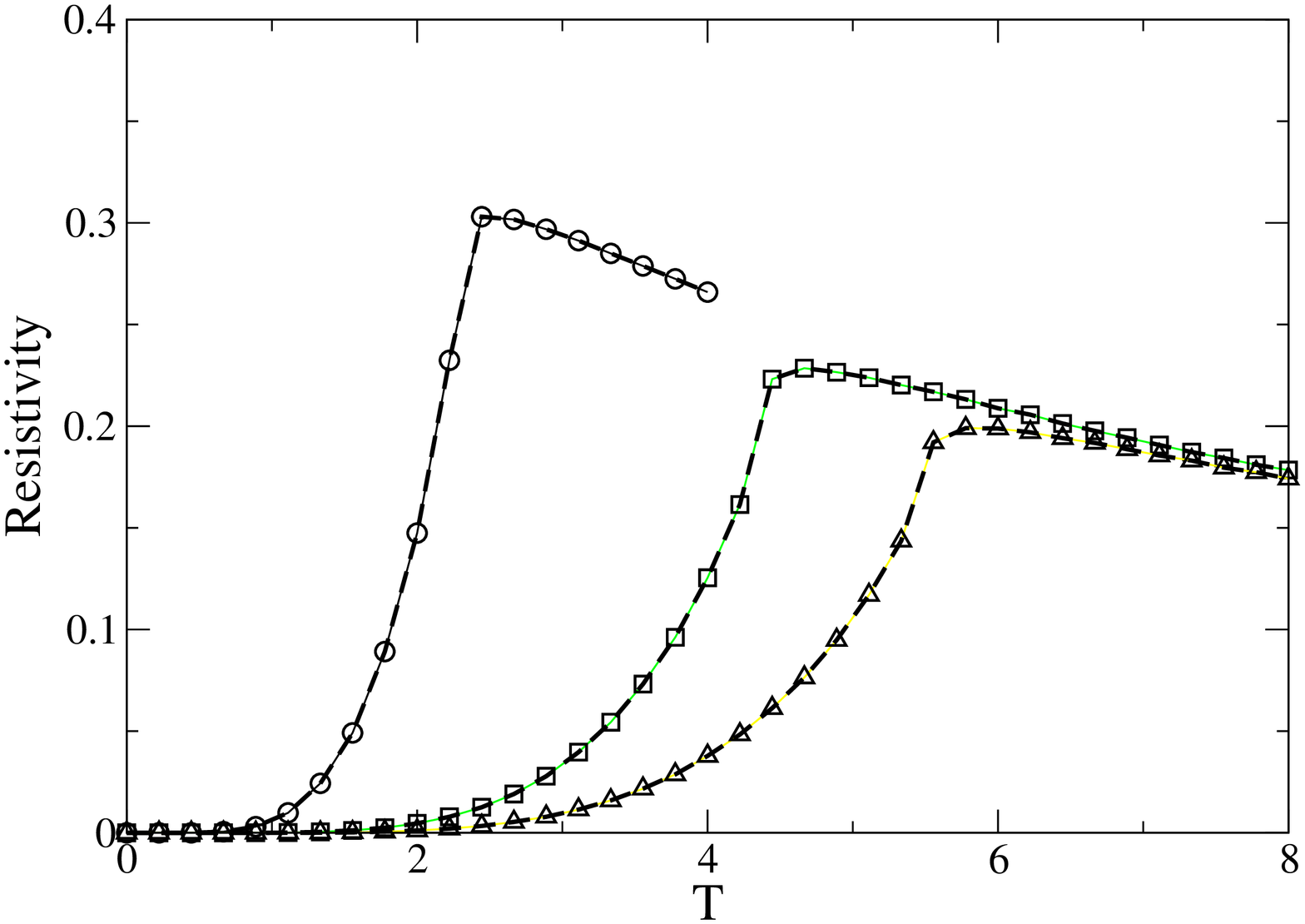}
 \includegraphics[width=70mm,angle=-90]{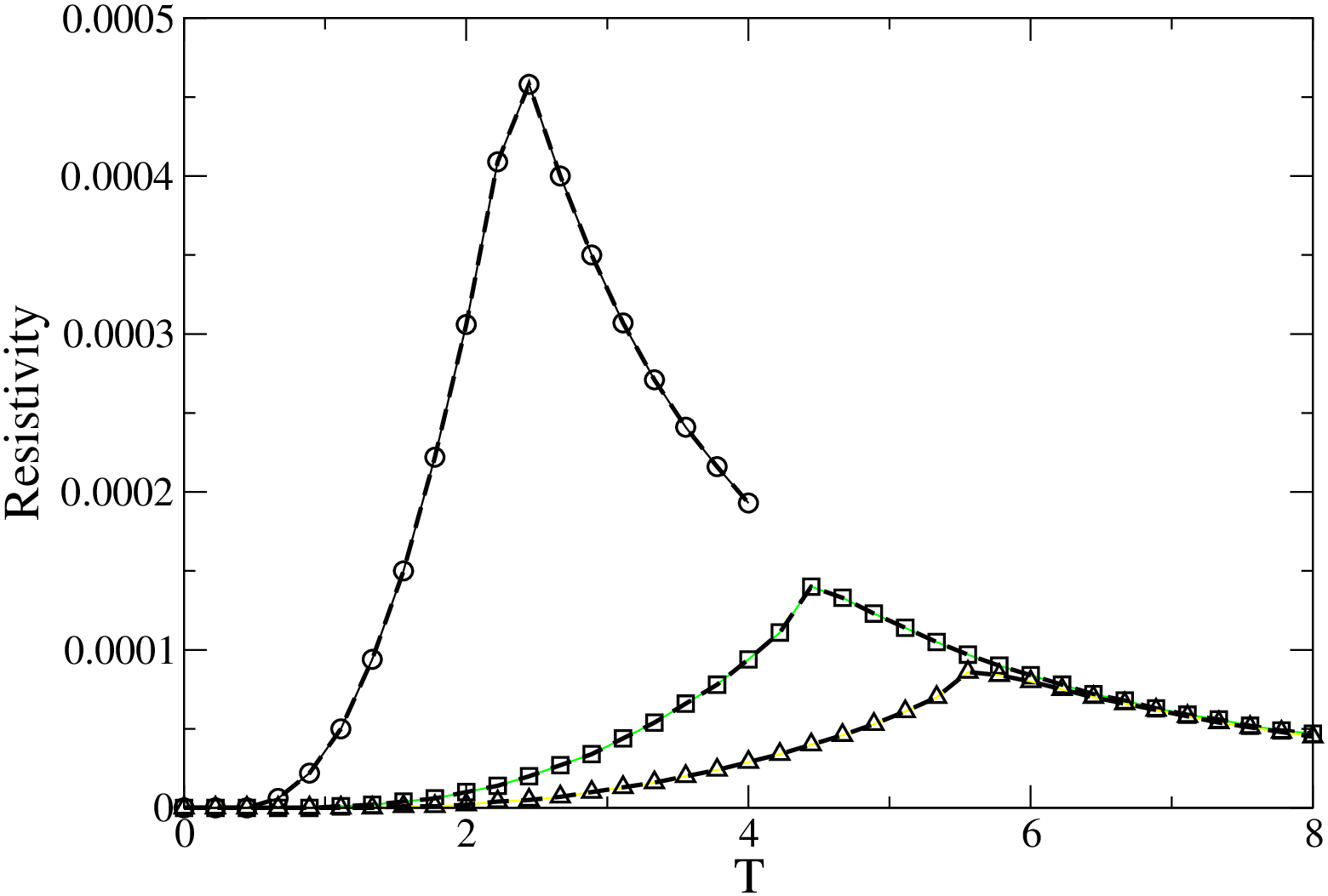}
 \caption{Spin resistivity versus $T$ for several anisotropy values $A$ in
 antiferromagnetic BCC system: $A=0.01$ (circles), 1 (squares), 1.5 (triangles). Upper (lower) curves:  degenerate (non degenerate) system. } \label{fig:RC}
\end{figure}

The results shown in Fig.~\ref{fig:RC} indicate clearly the appearance of a peak at the transition which
diminishes with increasing anisotropy. If we look at Fig.~\ref{cluster} which shows
the average size of clusters as a function of $T$, we observe that the size of clusters of large $S_z$
diminishes with increasing $A$.

\begin{figure}[h!]
 \centering
 \includegraphics[width=70mm,angle=-90]{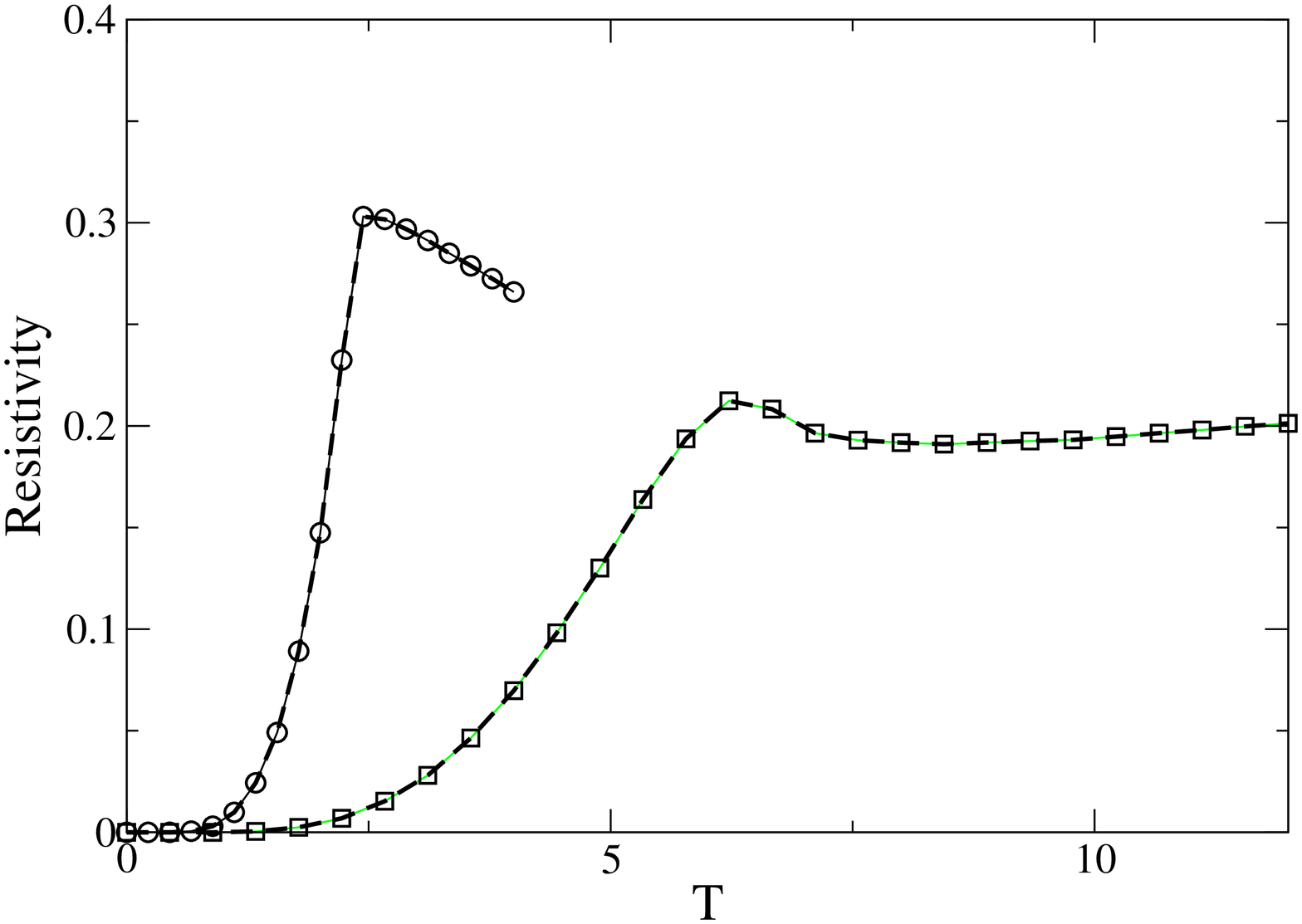}
 \includegraphics[width=70mm,angle=-90]{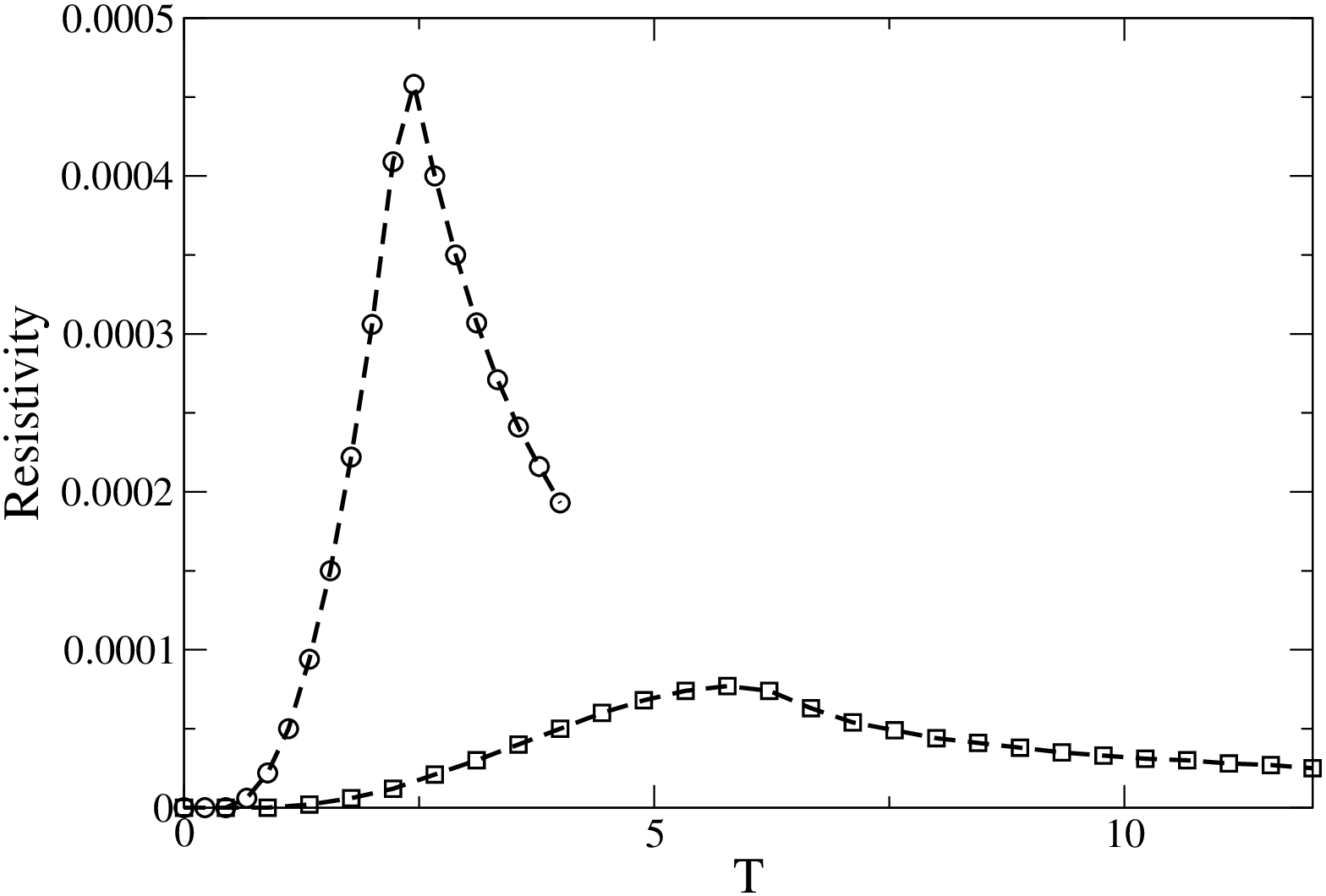}
 \caption{Spin resistivity for pure Heisenberg (circles) and Ising (squares) models in antiferroamgnetic BCC system. Upper (lower) curves:
 degenerate (non degenerate) system.} \label{fig:RI}
\end{figure}

We show in Fig.~\ref{fig:RI} the pure Heisenberg and Ising models.  For the pure Ising model, there is just a shoulder around $T_N$ with a different behavior in the paramagnetic phase: increase or decrease  with increasing $T$ for degenerate or non degenerate cases. It is worth to mention that MC simulations for the pure Ising model on the simple cubic and BCC antiferromagnets where interactions between itinerant spins are taken into account in addition to  Eq. (\ref{int}), show no peak at all\cite{Magnin,Magnin2}. These results are in agreement with the tendency observed here for increasing $A$.


\subsection{Effect of Magnetic Field}
We apply now  a magnetic field perpendicularly to the electric field. To see the effect of the magnetic
field it suffices to replace the distribution function by
\begin{equation}
f^1_k=\dfrac{e\hbar \tau_k}{m}(-\frac{\partial f^0}{\partial \epsilon})\textbf{k}.
\dfrac{(\textbf{E}-\dfrac{e\tau_k}{mc}\textbf{H}\wedge \textbf{E})}{1+(\dfrac{e\tau_kH}{mc})^2}
\end{equation}
From this, we obtain the following equations for the contributions of up and down spins
\begin{equation}
\rho_{\downarrow}=\sum^{+1}_{S_z=-1} (S_z+1)^2 \dfrac{\eta 4V_0^2 m^2 \pi k_f \xi^2}{n e^2 \hbar^3}[\dfrac{4\xi^2}{1+4\xi^2 k^2_f}]^2
\end{equation}

\begin{equation}
\rho_{\uparrow}=\sum^{+1}_{S_z=-1} (S_z-1)^2 \dfrac{\eta 4V_0^2 m^2 \pi k_f \xi^2}{n e^2 \hbar^3}[\dfrac{4\xi^2}{1+4\xi^2 k^2_f}]^2
\end{equation}
where $S_z$ is the domain-wall spin (scattering centers) and $V_0$ is the coefficient of the exchange integral between an itinerant spin and a lattice spin [see Eq. (\ref{exchange})].

\begin{figure}[h!]
 \centering
 \includegraphics[width=70mm,angle=0]{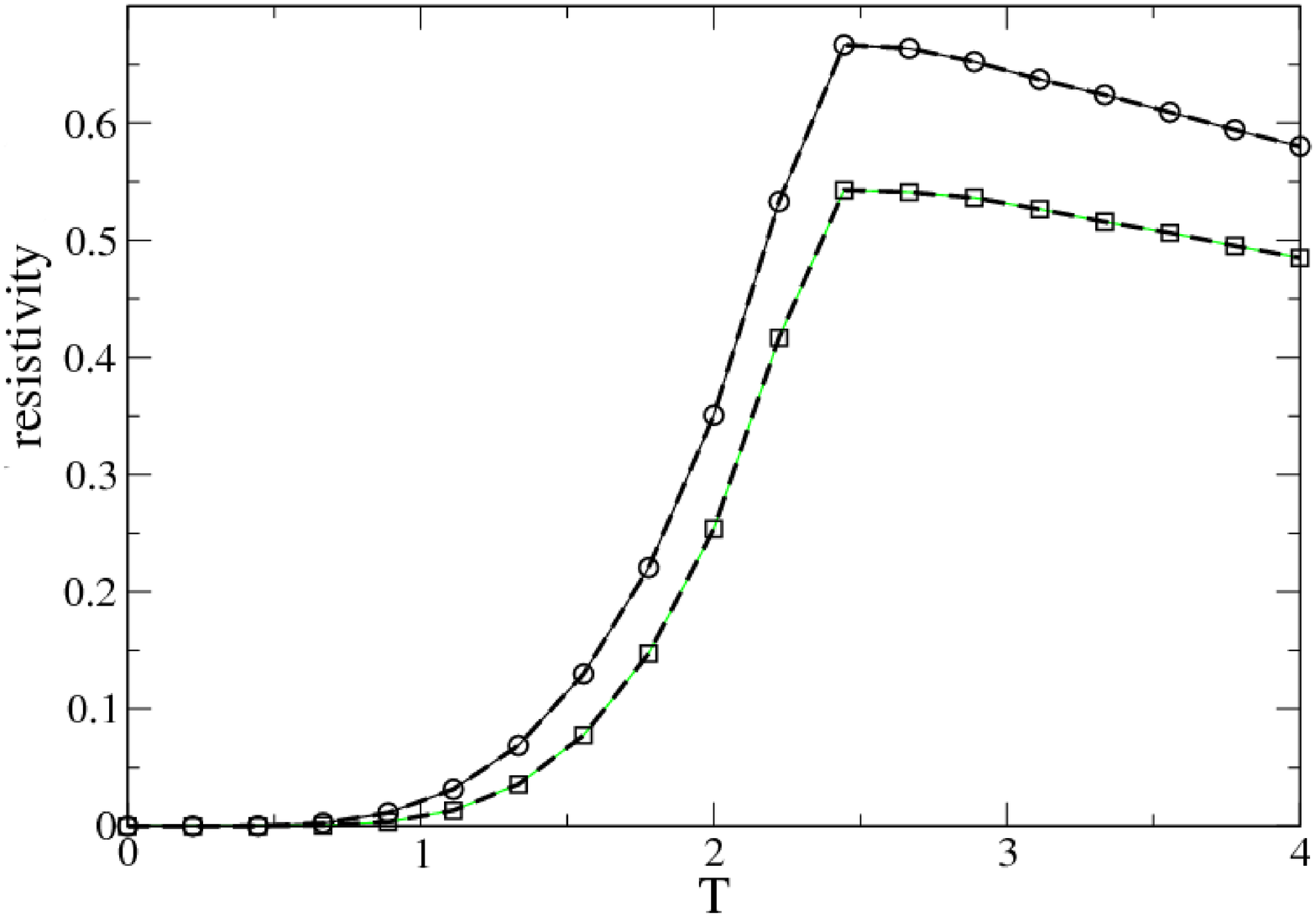}
 \includegraphics[width=70mm,angle=0]{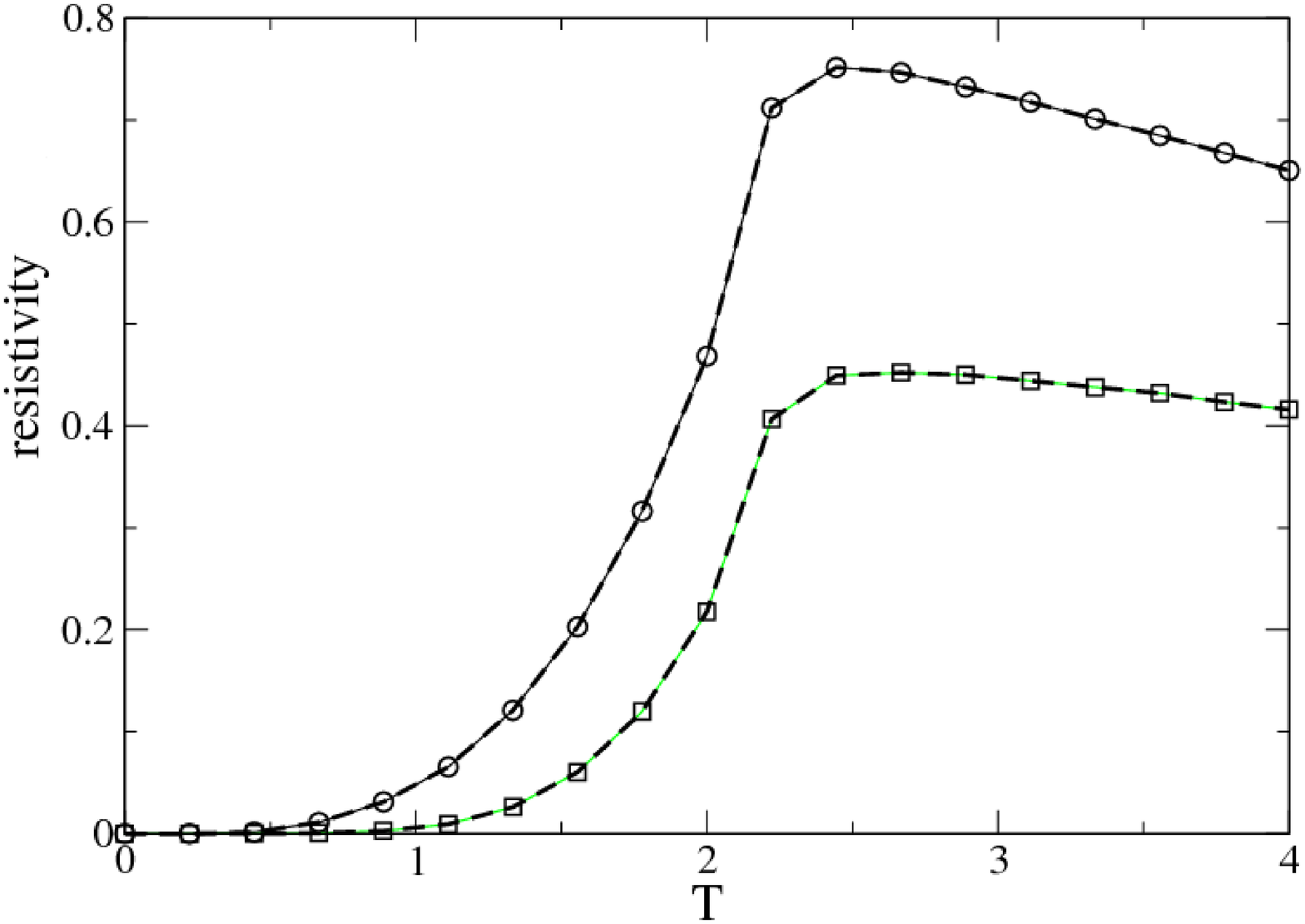}
  \caption{Resistivities of up (circles) and down (squares) spins versus $T$ for two magnetic field's strengths in the degenerate case. Top (bottom): $B=0.6 (1.5)$.} \label{fig:RB}
\end{figure}

\begin{figure}[h!]
 \centering
  \includegraphics[width=70mm,angle=0]{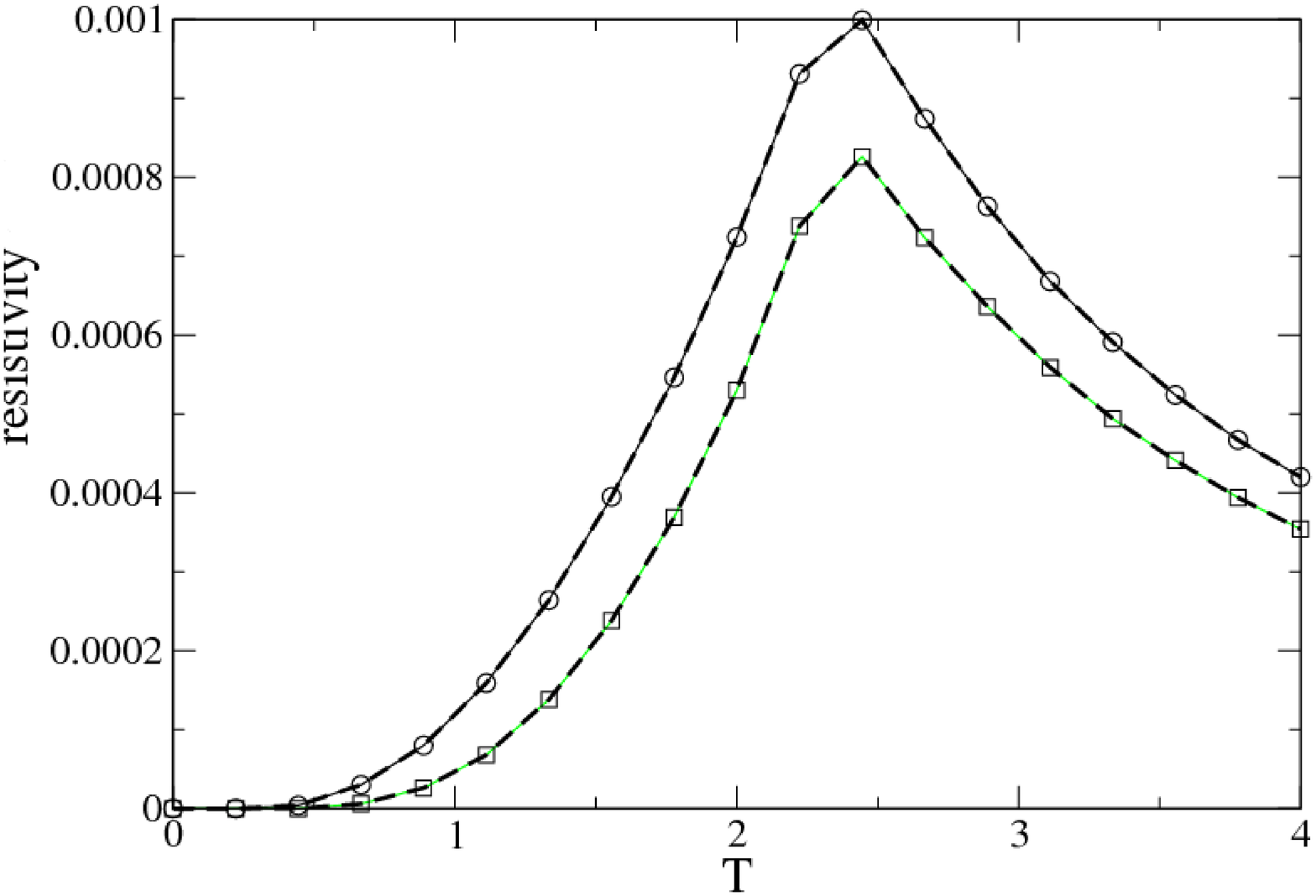}
 \includegraphics[width=70mm,angle=0]{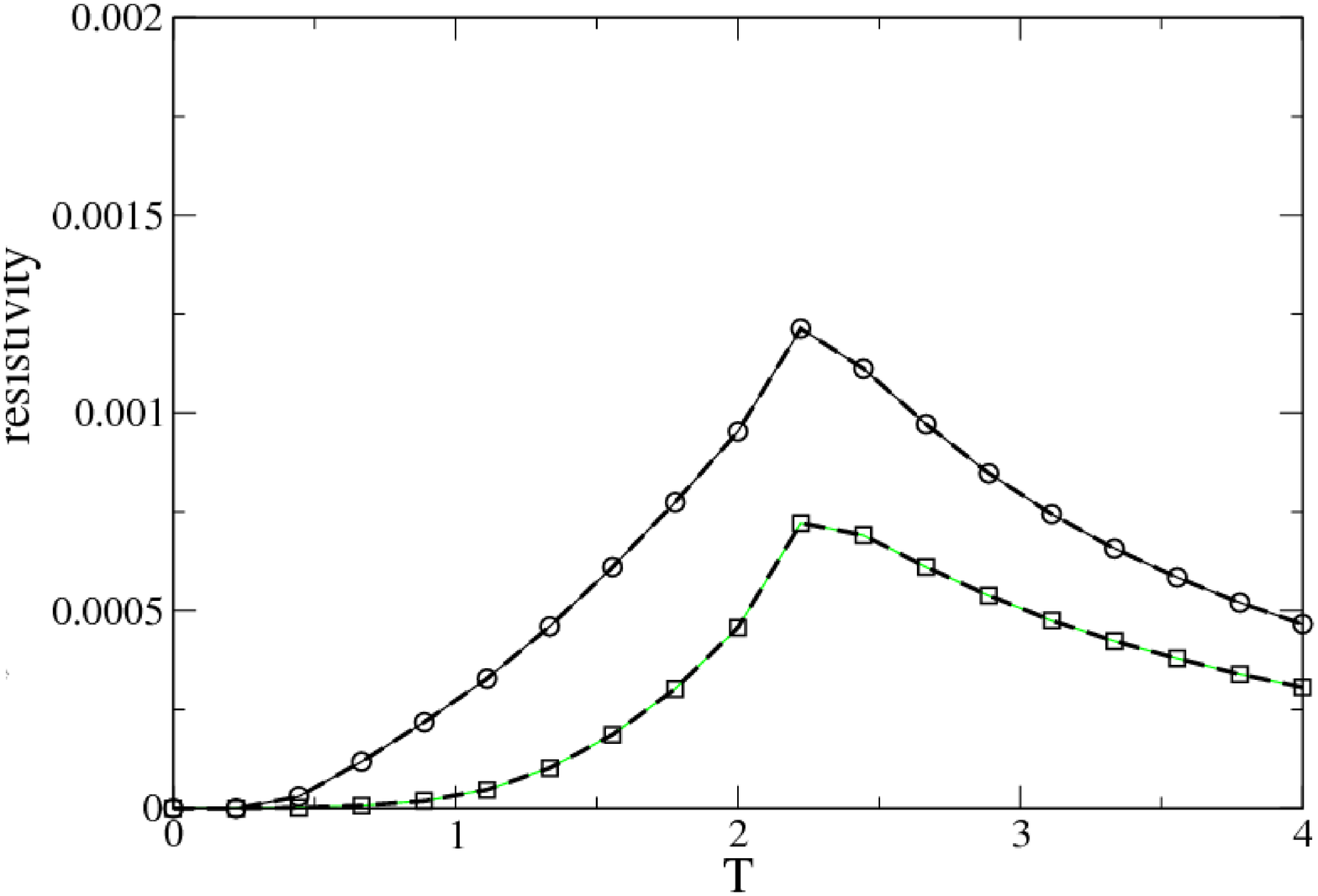}
 \caption{Resistivities of up (circles) and down (squares) spins versus $T$ for two magnetic field's strengths in the non degenerate case. Top (bottom): $B=0.6 (1.5)$.} \label{fig:RB1}
\end{figure}

\begin{figure}[h!]
 \centering
 \includegraphics[width=70mm,angle=-90]{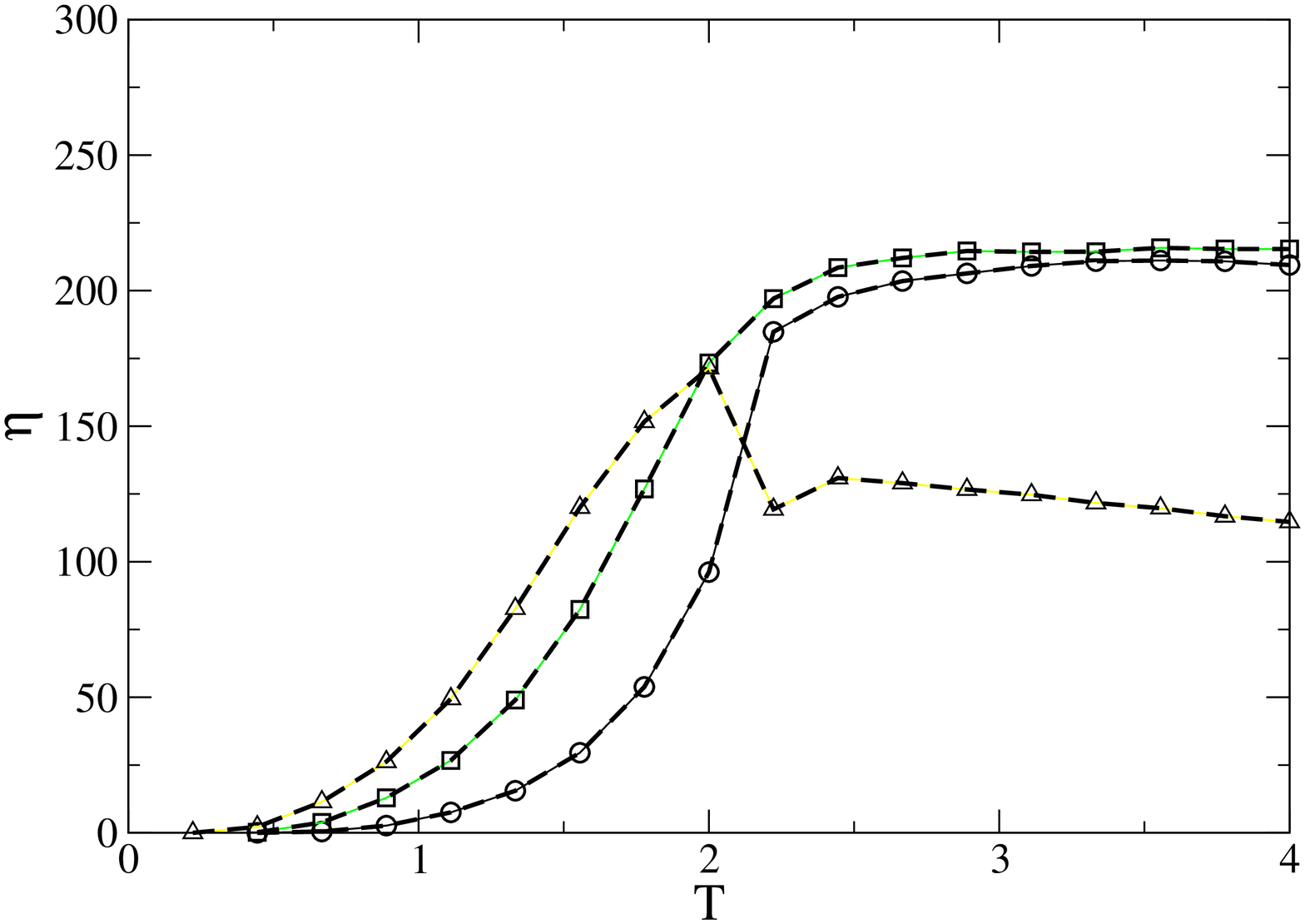}
 \includegraphics[width=70mm,angle=-90]{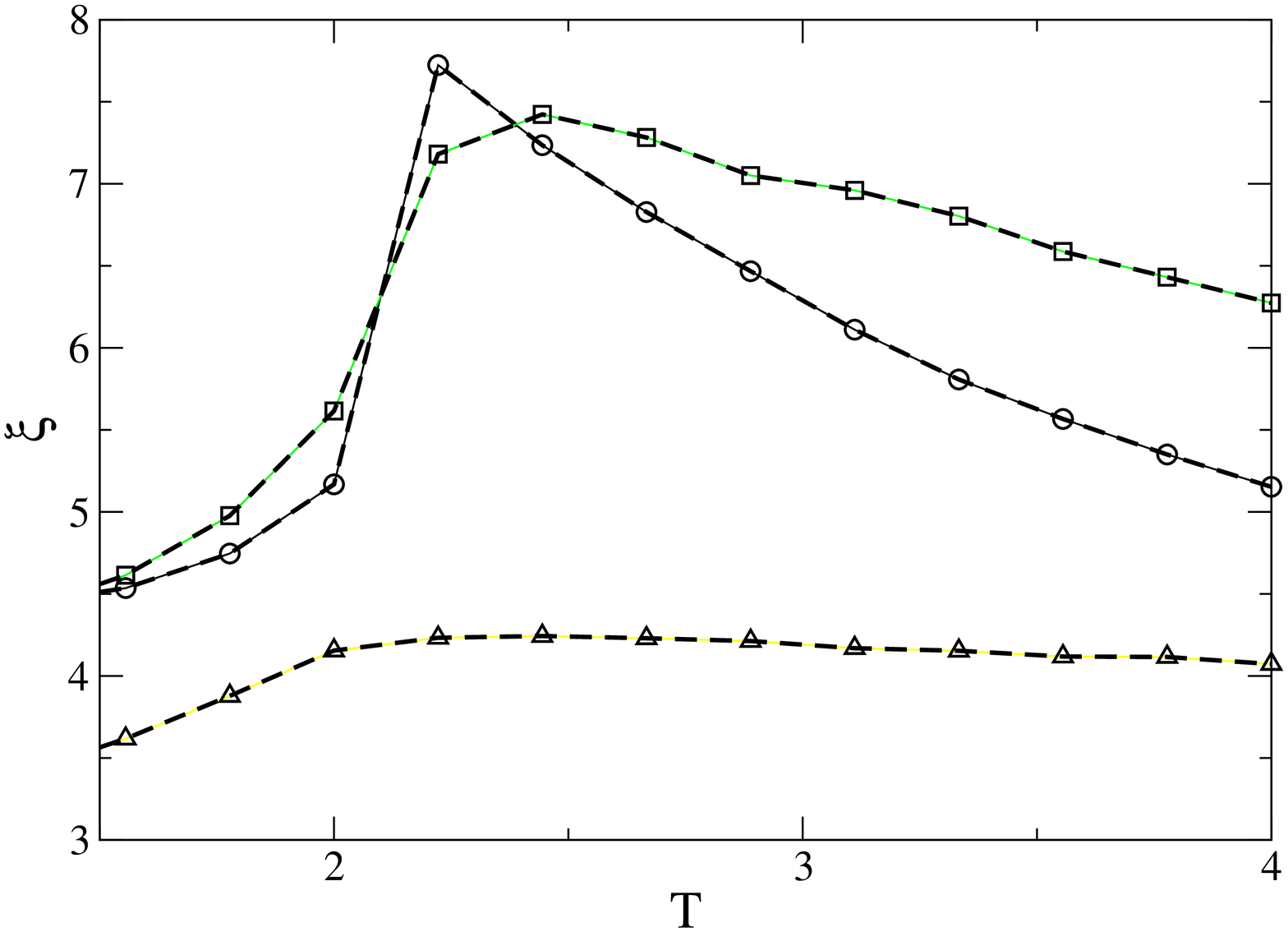}
\caption{Upper: Number of clusters, Lower: Average size of clusters,  versus $T$ for several values of $S_z$ and for magnetic field $B=1.5$.   Circles: $S_z=1$, squares: $S_z=0.8$, triangles: $S_z=0.6$. Lines are guides to the eye.} \label{AB}
\end{figure}

Figures~\ref{fig:RB} and \ref{fig:RB1} show the resistivity for several magnetic fields. We observe a split
in the resistivity for up and down spins which is larger for stronger field.  Also, we see
that the minority spins shows a smaller resistivity due to their smaller number. The reason is similar to the effect of $A$ mentioned above and
can be understood by examining Fig.~\ref{AB} where we show the evolution of the number and the average size of clusters with the temperature
in a magnetic field. By comparing with the zero-field results shown in Figs. \ref{histo} and \ref{cluster}, we can see that while the number of clusters does not change with the applied field, the size of clusters is significantly bigger. It is easy to understand
this situation: when we apply a magnetic field, the spins want to align themselves to the field so the up-spin domains become larger,  critical fluctuations are at least partially suppressed, the transition is softened.

\subsection{Application to MnTe}

We have chosen a presentation of the general model which can be applied to degenerate and non-degenerate semiconductors and semi metals.  The application to hexagonal MnTe is made below with the formulae of both degenerate and non-degenerate cases, for comparison.  Hexagonal MnTe has a big gap (1.27 eV), but it is an indirect gap. So, thermal excitations of electrons to the conduction band may not need to cross the gap channel. This may justify the use of the degenerate formulae.  In the degenerate case, $k_f$ depends only on the carrier concentration $n$ via the known formula:  $k_f=(3\pi^2n)^{1/3}$ .  We use for MnTe $n=2\times 10^{22}$cm$^{-3}$ mentioned below.
For the non-degenerate  case, $k_f$ is not necessary. Note that in the case of pure intrinsic semiconductors, $k_f$ is in the gap and its position is given by the law of mass action using parabolic band approximation.  In doped cases, band tails created by doped impurities can cover more or less the gap.   But this system, which is disordered by doping, is not a purpose of our present study.

In semiconductors  valence electrons can go from the valence band to the conduction band more and more as the temperature increases.  Therefore, the carrier concentration is a function of  $T$.
Our model has a number of itinerant spins which is independent of $T$ in each simulation.    However in each  simulation, we can take another concentration (see Ref. 19): the results show that the resistivity is not strongly modified, one still has the same feature, except that the stronger the concentration is the smaller the peak at $T_C$ becomes if and only if interaction between itinerant spins is taken into account. Therefore, we believe that generic effects independent of carrier concentration will remain.  Of course, the correct way is to use a formula to generate the carrier concentration as a function of $T$ and to make the simulation with the temperature-dependent concentration taking account additional scattering due to interaction between itinerant spins.  Unfortunately, to obtain that formula we have to use  several approximations which involve more parameters.  We will try this in a future work.

In the case of Cd$_{1-x}$Mn$_x$Te, the question of the crystal structure, depending
on the doping concentration $x$ remains open.   Cd$_{1-x}$Mn$_x$Te can have one of the following structures, the so-called
NiAs structure or the zinc-blend one, or a mixed phase.\cite{Komatsubara,Szwacki,Wei,Adachi}

\begin{figure}[h!]
 \centering
 \includegraphics[width=60mm,angle=0]{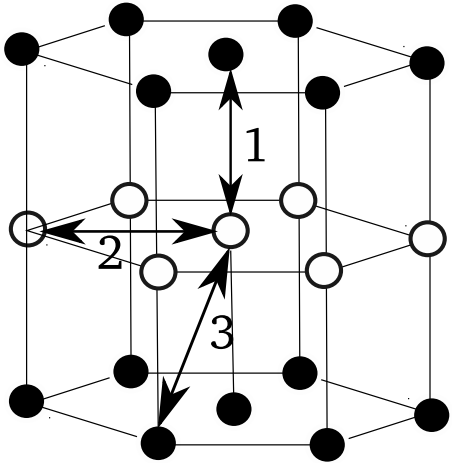}
 \caption{Structure of the type NiAs is shown with  Mn atoms only.  This is a stacked hexagonal lattice. Up spins are shown by black circles, down spins by white ones.  Nearest-neighbor (NN) bond is marked by 1, next NN bond by 2, and third NN bond by 3.} \label{NiAs}
\end{figure}

The pure MnTe
crystallizes in either the zinc-blend structure\cite{Hennion} or the hexagonal NiAs one\cite{Hennion2} (see Fig.~\ref{NiAs}).  MnTe is a well-studied $p$-type semiconductor with numerous applications due to its high N\'eel temperature.
We are interested here in the case of hexagonal structure. For this case, the N\'eel temperature is $T_N=310$ K\cite{Hennion2}.

The cell parameters are $a=4.158 \AA$ and $c=6.71 \AA$ and we have an indirect band gap of $E_g=1.27 $eV.

Magnetic properties are determined mainly by an antiferromagnetic exchange integral
between nearest-neighbors (NN) Mn along the $c$ axis, namely  $J_1/k_B=-21.5\pm 0.3$ K,  and a ferromagnetic exchange
$J_2/k_B \approx 0.67\pm 0.05$ between in-plane  (next NN) Mn.  Third NN interaction has been also measured with $J_3/k_B\simeq -2.87\pm 0.04$ K.  Note that the spins are lying in the $xy$ planes perpendicular to the $c$ direction with an in-plane easy-axis anisotropy\cite{Hennion2}.
The magnetic structure is therefore composed of ferromagnetic $xy$ hexagonal planes antiferromagnetically stacked in the $c$ direction.  The NN distance in the $c$ direction is therefore $c/2\simeq 3.36$ shorter than the in-plane NN distance $a$.

\begin{figure}[h!]
 \centering
 \includegraphics[width=70mm,angle=-90]{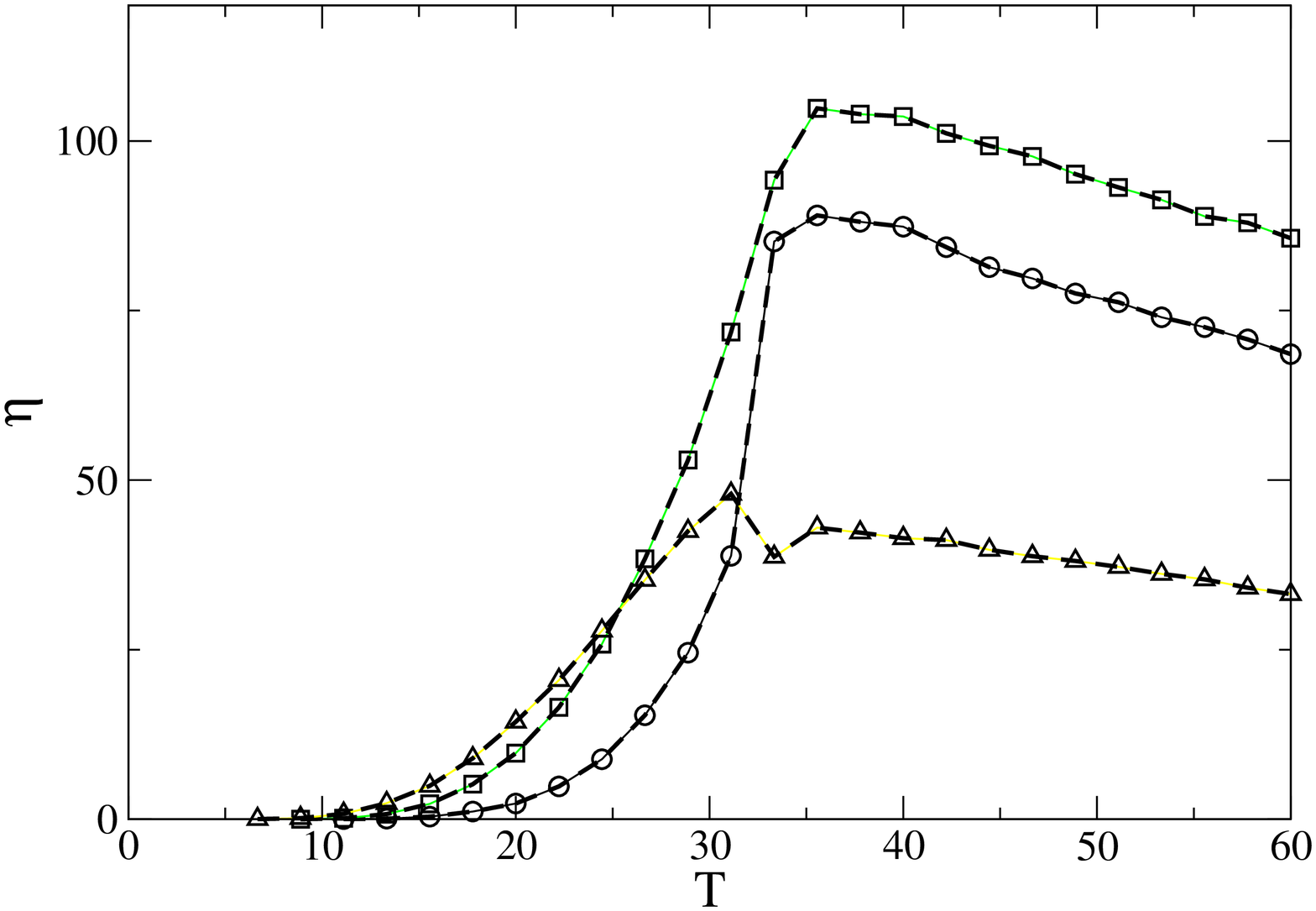}
 \includegraphics[width=70mm,angle=-90]{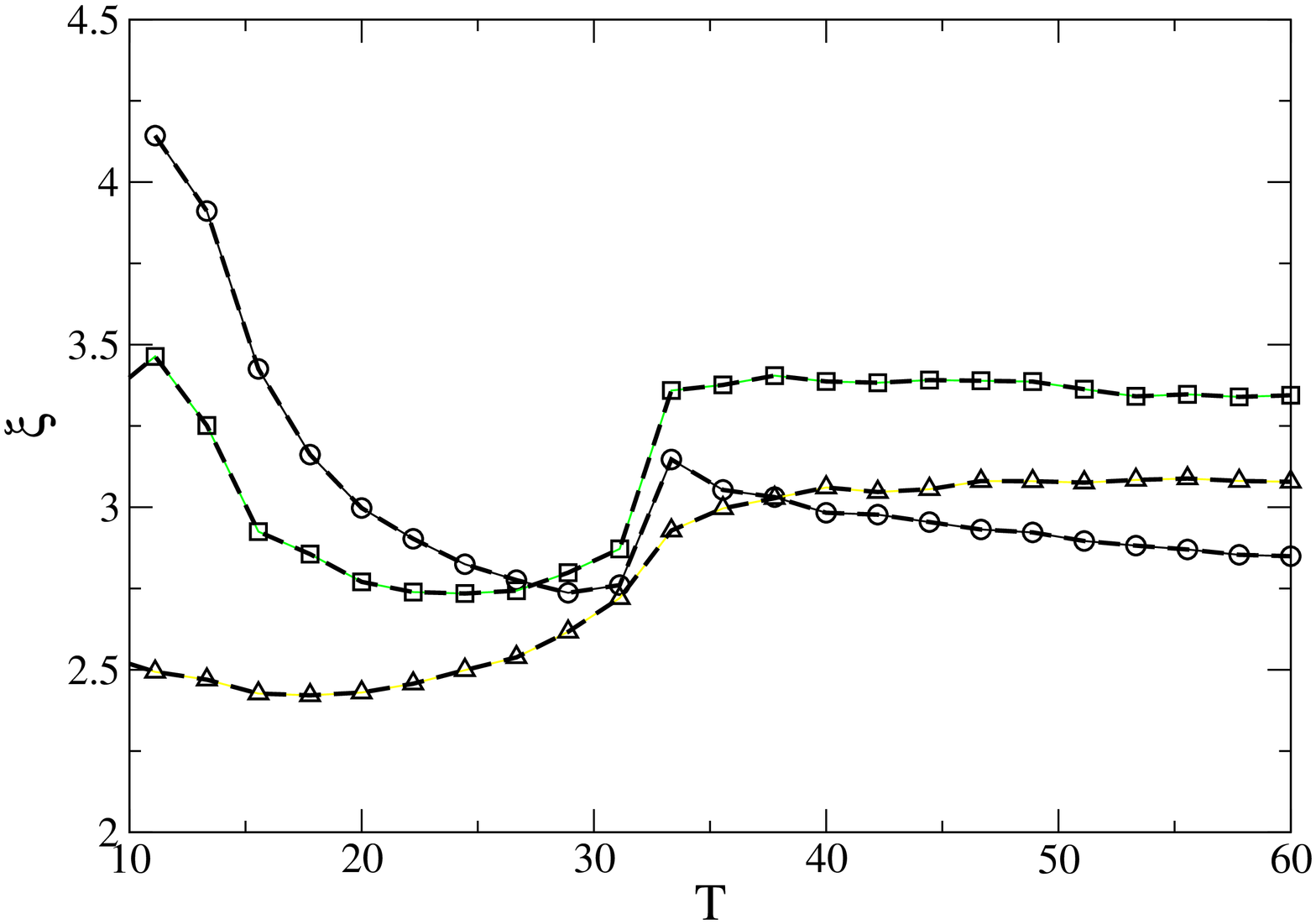}
\caption{Number of clusters (upper) and cluster size (lower) versus  $T$ for MnTe structure obtained from Monte Carlo simulations for several values of $S_z$: 1 (circles), 0.8 (squares), 0.6 (triangles).  Lines are guides to the eye.} \label{NMnTe}
\end{figure}

\begin{figure}[h!]
 \centering
 \includegraphics[width=70mm,angle=-90]{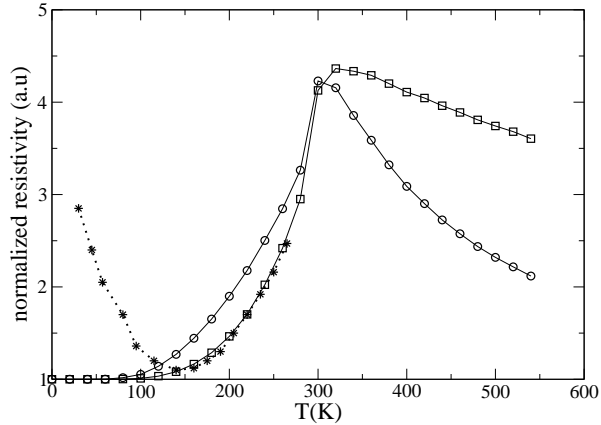}
 \caption{Normalized spin resistivity versus $T$ in MnTe : theoretical non-degenerate case (circles), theoretical degenerate case (squares)
and experimental results (stars) from Chandra
 et al\cite{Chandra}.  Experimental data lie on the degenerate line for $T\geq 140$ K. See text for comments.} \label{fig:MnTe}
\end{figure}

We have calculated the cluster distribution for the hexagonal MnTe using the exchange integrals taken from Ref. \cite{Hennion2} and the other crystal parameters  taken from the literature\cite{Inoue,Okada,Chandra}. The result is shown in
Fig.~\ref{NMnTe}.  The spin resistivity in MnTe obtained with our theoretical model is presented  in
Fig.~\ref{fig:MnTe} for a density of  itinerant spins corresponding to $n=2\times 10^{22}$ cm$^{-3}$,  together with "normalized" experimental data.  The normalization has been made by noting that the experimental resistivity $R$ in  Ref. 41 is the total one with contributions from impurities and phonons. However, the phonon contribution is important only at high $T$, so we can neglect it for $T<310$K. While for the contribution $R_0$ from fixed impurities, there are reasons to consider it as  temperature-independent at low $T$.  From these rather rude considerations, we extract  $R_0$ from $R$ and compare our theoretical with $R-R_0$.  This is what we called "normalized resistivity" in Fig.~\ref{fig:MnTe}.

Several remarks are in order:

i) the peak temperature of our theoretical model is found at 310 K corresponding the the experimental N\'eel temperature although for our fit we have used only the above-mentioned values of exchange integrals

ii) our result is in agreement  with experimental data obtained by Chandra et al.\cite{Chandra}  for temperatures between 140 K and 280 K above which Chandra et al. did not unfortunately measured

iii) at temperatures lower than 140 K, the experimental curve increases with decreasing $T$. Note that many experimental data on various materials show this 'universal' feature: we can mention the data by Li et al.\cite{Li}, Du et al.\cite{Du}, Zhang et al.\cite{Zhang}, McGuire et al.\cite{McGuire} among others. 
Our theoretical model based on the scattering by defect clusters cannot account for this behavior because there are no defects at very low $T$.  Direct Monte Carlo simulation shows however that the freezing indeed occurs at low $T$ both in ferromagnets\cite{Magnin,Akabli3} and antiferromagnets\cite{Magnin2} giving rise to an increase of the spin resistivity with decreasing $T$. There are several explanations for this experimental behavior among which 
we can mention the fact that in semiconductors the carrier concentration increases as $T$ increases, giving rise to an increase of the spin current, namely a decrease of the resistivity, with increasing $T$ in the low-$T$ region.
Another origin of the increase of $\rho$ as $T\rightarrow 0$ is the possibility that
the itinerant electrons may be frozen (crystallized) due to their interactions with localized spins and between themselves, giving rise to low mobility.  
On the hypothesis of frozen electrons, there is a reference on the charge-ordering at low $T$ in Pr$_{0.5}$Ca$_{0.5}$MnO$_3$\cite{Zhang} due to some strain interaction. A magnetic field can make this ordering melted giving rise to a depressed resistivity.  Our present model does not correspond to this compound but we believe that the concept is similar.  For the system
Pr$_{0.5}$Ca$_{0.5}$MnO$_3$, which shows a commensurate charge order, the "melting"
fields at low temperatures are high, on the order of 25 Tesla\cite{Zhang}.

iv) the existence of the peak at $T_N=310$ K of the theoretical spin resistivity shown in Fig.~\ref{fig:MnTe} is in agreement with experimental data recently published by Li et al.\cite{Li} (see the inset of their Fig. 5).    Unfortunately, we could not renormalize the resistivity values of Li et al.\cite{Li} to put in the same figure with our result for a quantitative comparison.  Other data on various materials\cite{Du,McGuire,Zhang} also show a large peak at the magnetic transition temperature.

To close this section, let us note that it is also possible, with some precaution, to apply our model on other families of antiferromagnetic semiconductors
 like CeRhIn$_5$ and LaFeAsO. An example of supplementary difficulties but exciting subject encountered in the latter compound is that there are two transitions in a small temperature region:  a magnetic transition at $145$ K and a tetragonal-orthorhombic
 crystallographic phase transition at $160$ K.\cite{McGuire,Christianson}  An application to  ferromagnetic semiconductors of the n-type CdCr$_2$Se$_4$\cite{Lehmann2} is under way.

 \section{Conclusion}

 We have shown in this paper the behavior of the magnetic resistivity
 $\rho$ as a function of temperature in antiferromagnetic semiconductors.
 The main interaction which governs the resistivity behavior  is the interaction
 between itinerant spins and the lattice spins.  Our analysis, based on the Boltzmann's equation which
 uses the temperature-dependent cluster distribution obtained by MC simulation.  Our result
 is in agreement with the theory by Haas\cite{Haas}: we observe a broad maximum of $\rho$ in the
 temperature region
 of the magnetic transition without a sharp peak observed in ferromagnetic materials.  We have studied the two cases, degenerate and non-degenerate semiconductors.
 The non-degenerate case shows a maximum which is more pronounced than
 that of the degenerate case.  We would like to emphasize that the shape of the maximum and  its existence depend on several physical parameters such as interactions between different kinds of spins, the spin model, the crystal structure etc.
 In this paper we applied our theoretical model in the antiferromagnetic semiconductor MnTe.  We found a good agreement with experimental data near the transition region.  We note however that our model using the cluster distribution cannot be applied at very low $T$ where the spin resistivity in experiments is dominated by effects other than $s-d$ scattering model of the present paper.  One of these possible effects is the carrier proliferation with increasing  temperatures in semiconductors which makes the resistivity decrease with increasing $T$ experimentally observed in magnetic semiconductors at low $T$.

\section*{Ackowledgments}

 One of us (KA) wishes to thank the JSPS for a financial support
 of his stay at Okayama University where this work was carried out.
 He is also grateful to the researchers of Prof. Isao Harada's group for helpful discussion.

{}

\end{document}